# Common kernel-smoothed proper orthogonal decomposition (CKSPOD): An efficient reduced-order model for emulation of spatiotemporally evolving flow dynamics

Yu-Hung Chang[a], Xingjian Wang[a,b], Liwei Zhang,[a,c] Yixing Li[a], Simon Mak[d,e], Chien-Fu J. Wu[d], Vigor Yang[a,*,¶]

[a]*School of Aerospace Engineering, Georgia Institute of Technology, Atlanta, Georgia, USA*
[b]*Mechanical and Civil Engineering, Florida Institute of Technology, Melbourne, Florida, USA*
[c]*Department of Mechanical & Aerospace Engineering, University of Texas Arlington, Arlington, Texas, USA*
[d]*School of Industrial and Systems Engineering, Georgia Institute of Technology, Atlanta, Georgia, USA*
[e]*Department of Statistical Science, Duke University, Durham, North Carolina, USA*

**In the present study, we propose a new surrogate model – common kernel-smoothed proper orthogonal decomposition (CKSPOD) – to efficiently emulate the spatiotemporal evolution of fluid flow dynamics. The proposed surrogate model integrates and extends recent developments in Gaussian process learning, high-fidelity simulations, projection-based model reduction, uncertainty quantification, and experimental design, rendering a systematic, multidisciplinary framework. The novelty of the CKSPOD emulation lies in the construction of a common Gram matrix, which results from the Hadamard product of Gram matrices of all observed design settings. The Gram matrix is a spatially averaged temporal correlation matrix and contains the temporal dynamics of the corresponding sampling point. The common Gram matrix synthesizes the temporal dynamics by transferring POD modes into spatial functions at each observed design setting, which remedies the phase-difference issue encountered in the kernel-smoothed POD (KSPOD) emulation – a recent**

\* Corresponding author (Email: vigor.yang@aerospace.gatech.edu)

**fluid flow emulator proposed in Chang et al. (2020). The CKSPOD methodology is demonstrated through a model study of flow dynamics of swirl injectors with three design parameters. A total of 30 training design settings and 8 validation design settings are included. Both qualitative and quantitative results show that the CKSPOD emulation outperforms the KSPOD emulation for all validation cases, and is capable of capturing small-scale wave structures on the liquid-film surface faithfully. The turbulent kinetic energy prediction using CKSPOD reveals lower predictive uncertainty than KSPOD, thereby allowing for more accurate and precise flow predictions. The turnaround time of the CKSPOD emulation is about 5 orders of magnitude faster than the corresponding high-fidelity simulation, which enables an efficient and scalable framework for design exploration and optimization.**

**fluid flow emulator proposed in Chang et al. (2020). The CKSPOD methodology is demonstrated through a model study of flow dynamics of swirl injectors with three design parameters. A total of 30 training design settings and 8 validation design settings are included. Both qualitative and quantitative results show that the CKSPOD emulation outperforms the KSPOD emulation for all validation cases, and is capable of capturing small-scale wave structures on the liquid-film surface faithfully. The turbulent kinetic energy prediction using CKSPOD reveals lower predictive uncertainty than KSPOD, thereby allowing for more accurate and precise flow predictions. The turnaround time of the CKSPOD emulation is about 5 orders of magnitude faster than the corresponding high-fidelity simulation, which enables an efficient and scalable framework for design exploration and optimization.**

**Nomenclature**

| | | |
|---|---|---|
| $\alpha$ | = | liquid film spreading angle at injector exit |
| $\beta$ | = | time-varying coefficients |
| $C$ | = | covariance matrix |
| $d$ | = | design point (parameter set) |
| $\delta$ | = | inlet slot width |
| $\Delta L$ | = | distance between injector inlet and headend |
| $f$ | = | flow property |
| $f_h$ | = | probability density function |
| $h$ | = | liquid film thickness at injector exit |
| $i, j, k$ | = | dummy indices |
| K | = | geometric constant |
| $L$ | = | injector length |
| $\mu$ | = | mean |
| $p$ | = | design parameters |
| $\phi$ | = | spatial basis functions |
| $\boldsymbol{R}$ | = | Gaussian correlation function |
| $R_n$ | = | injector radius |
| $\rho$ | = | density, kg/m³ |
| $\boldsymbol{\Lambda}$ | = | diagonal matrix |
| $\boldsymbol{T}$ | = | temperature, K |
| $\boldsymbol{\mathcal{T}}$ | = | CKSPOD transfer matrix |
| $t$ | = | time |
| $\theta$ | = | tangential inlet angle |
| $\boldsymbol{U}$ | = | left-singular vectors by SVD |
| $\boldsymbol{V}$ | = | right-singular vectors by SVD |
| $w$ | = | kriging weighing number |
| $x$ | = | spatial coordinate |

# 1. Introduction

In the study of complex natural phenomena and engineering systems, high-fidelity simulations have been used for decades to provide detailed quantification of flow structures and dynamics that are otherwise hard to obtain from experiments or theoretical analyses [1-7]. These simulations, however, are computationally expensive and time consuming. For example, the simulation of propellant mixing in an azimuthal sector of a bi-swirl injector with 1.6 M cells takes 10,000 CPU hours (hexa-core AMD Opteron Processor 8431) to sweep 1 millisecond of flow time using large-eddy simulation (LES) [6]. It is thus impractical to rely solely on high-fidelity simulations for practical design, which often requires the survey of a wide parametric space. Surrogate models have been implemented to improve design efficiency. The objective of this study is to develop and validate an efficient surrogate model (emulator) to predict a spatiotemporally evolving flowfield in a faithful manner.

Surrogate models are designed to closely mimic the results of high-fidelity simulation with significantly reduced computing time and resource [8-11]. Surrogate models can be broadly categorized into data-fit, multifidelity (hierarchical), and reduced-order models (ROMs) [12]. Data-fit models are not physics-based and are formulated directly from interpolation or regression of simulation data. Such models fit a response surface connecting input and output data, by evaluating objective and constraint functions in the sampling space. Jones [13] presented a taxonomy of response surface models (RSMs) using a general function to represent different data-fit models, including polynomial response surface models, kriging [14-17], and radial basis functions (RBFs) [18, 19]. Another data-fit model is support vector regression [9], which can be viewed as an extension of RBFs; this model designates a threshold within which regression error in the sample data is acceptable, without affecting surrogate prediction. Data-fit surrogate models have been widely used in aerospace system design and optimization, but these models can face stiff challenges in problems with dynamic evolutions and high-dimensional data. Multifidelity (hierarchical) models can alleviate such an issue, by incorporating low-fidelity models into high-fidelity models with a bridge function [20-22], where low-fidelity models are established via simplification of physical assumptions or reduction of numerical resolution. However, the prediction accuracy of multifidelity models may be compromised when the low-fidelity models provide a completely different trend from the high-fidelity counterpart.

ROMs are also frequently used to overcome the curse of dimensionality issue by constructing a low-dimensional subspace, on which the reduced operators or basis functions are sought from the original high-dimensional datasets. Projection-based ROMs are currently the most popular ROMs. This approach is often physics-based, since the reduced subspace typically contains the key structures of the dynamical system. Projection-based ROMs are particularly effective for systems whose input-output map is of low rank [23, 24]. The most widely used projection-based ROM in engineering is the *proper orthogonal decomposition* (POD) [25]. POD is inherently connected to the idea of principal component analysis in the area of statistical learning [26] and Karhunen–Loève decomposition in the stochastic process literature [27]. In many problems, the POD-based ROM can capture key structures and dynamics embedded in high-fidelity data, which can then be used for prediction of a spatiotemporally evolving flowfield.

The use of POD-based ROMs has been discussed in many studies. In an early attempt, Ly et al. [28] applied POD to study the temperature field in a Rayleigh–Bénard convection problem, and used a cubic spline interpolation to predict the POD coefficients. Audouze et al. [29] employed radial basis functions to model POD coefficients, and validated the result against steady-state convection-diffusion-reaction problems. Swischuk et al. [30] proposed a physics-based parametric surrogate model using POD. In this paper, several machine learning methods, including neural networks, multivariate polynomial regression, k-nearest-neighbors, and decision trees, were used to learn the map between input parameters and POD expansion coefficients. They considered two engineering examples, and found that embedded physical constraints were important for the learned models. While these works have introduced some novel ideas for projection-based models, there are considerable limitations. First, these works considered only spatial or temporal development; no effort was given to problems with concurrent spatial and temporal evolution. Further, all the methods proposed thus far have focused on training POD coefficients, with little attention to POD modes; this may be problematic when applied to cases with complicated dynamics and varying flow conditions and geometric parameters. The POD-based ROM has also been questioned for its intrinsically linear subspaces, although some nonlinearity may be embedded in POD expansion coefficients. Dynamic mode decomposition (DMD) [31, 32] was developed to represent nonlinear, finite-dimensional dynamics without linearization by approximating the modes of the Koopman operator.

In recent years, deep learning methods, in particular, autoencoders, have been attempted to construct ROMs for engineering applications, due to their ability to treat system nonlinearity [33-37]. Autoencoders employ a neural-network structure with two elements: an encoder for the nonlinear mapping from the high-dimensional input to low-dimensional manifolds, and a decoder for nonlinear mapping from low-dimensional manifolds to an approximate representation of the high-dimensional input. Lee and Carlberg [35] implemented deep convolutional autoencoders to obtain reduced nonlinear manifolds of dynamical systems, which outperformed the linear subspace ROMs in selected advection-dominated problems. Xu and Duraisamy [37] proposed a three-level convolutional autoencoder network, including a convolutional autoencoder, a temporal convolutional autoencoder, and a fully-connected network, for parametric and future-state predictions of spatiotemporally evolving systems. Although deep learning methods embed nonlinear features through activation functions, the physical connection between reduced nonlinear manifolds trained by deep learning and actual dynamical structures remains unclear. POD-based ROMs, on the other hand, provide a direct link between POD spatial modes and coherent structures of turbulent flows [25].

In this paper, we focus on the development of POD-based surrogate models, primarily due to their valuable physical interpretation of flow dynamics. For situations with different geometries, a common-grid POD (CPOD) technique was recently established, capable of handling the spatiotemporal evolution of the flowfield at various design points [17, 38]. As a demonstration case, the method was applied to study the flow dynamics of swirl injectors. The mean flow structures were successfully predicted over a broad range of geometric dimensions, but the accuracy of the prediction of instantaneous flowfields required further improvement. A kernel-smoothed POD (KSPOD) technique was then developed [39]. It employs kriging-based weighing functions for all the POD modes in the design matrix to construct spatial functions at a new design point. The KSPOD surrogate model significantly improves the accuracy of prediction of a spatiotemporally evolving flowfield, but one limitation is that its predictive accuracy drops if the training samples exhibit distinct physics at different design points. This has motivated the development of the proposed surrogate model, referred to as the *common kernel-smoothed POD* (CKSPOD) model. A common Gram matrix using the Hadamard product is established based on the Gram matrices at each sampling point to remedy the deficiency of the KSPOD technique when the training data exhibits distinct physics. Every Gram matrix contains the temporal dynamics of the corresponding sampling point, and the common Gram matrix synthesizes these dynamics through element-wise multiplication. The CKSPOD technique is able to efficiently predict complex flowfields over a broad range of operating conditions and geometric parameters [40], and can be efficiently incorporated into a data-driven emulation framework for design analysis and optimization [41].

The present paper is structured as follows. Section II provides a detailed description of the CKSPOD methodology, including the proposed surrogate modeling framework and its training algorithm. Section III discusses the uncertainty quantification associated with the CKSPOD model. In Section IV, the framework is applied for studying the spatiotemporal evolution of flow swirl injectors. Section V concludes this work with thoughts on future work.

## 2. Common kernel-smoothed POD (CKSPOD) surrogate modeling

### 2.1. The CKSPOD

In this section, we introduce a novel emulation technique, common kernel-smoothed proper orthogonal decomposition (CKSPOD). CKSPOD provides an efficient way to train a projection-based surrogate model using simulation results at observed design settings, and allows for predictions (emulations) over the desired design space in practical turn-around times. It also circumvents the issue of phase differences (sign differences of eigenvectors) commonly found in the KSPOD method by constructing a transfer matrix in the data reduction process.

The fundamentals of POD are briefly introduced first. POD is a model reduction method that extracts orthogonal basis functions and associated temporal coefficients based on the energy norms. From a physical perspective, POD provides valuable insight into the coherent structures in a flowfield. For a spatiotemporally evolving flow, the variable of interest $f$ at spatial location $x$ and time $t$, can be written as

$$f(x,t) = \sum_{k=1}^{m} \beta^k(t)\phi^k(x), \qquad (1)$$

where $\beta^k(t)$ and $\phi^k(x)$ represent the time-varying coefficient and spatial function (mode shape) of the $k$-th mode, respectively. $\phi^k(x)$ can be interpreted as the spatial distribution of the fluctuation field of a given flow variable (for example, pressure, density, temperature, and velocity components). The temporal coefficient $\beta^k(t)$ characterizes the dynamic evolution of the mode. A spectral analysis of $\beta^k(t)$ can be employed to identify flow periodicity and corresponding characteristic frequencies. The index $k$ in Eq. (1) denotes the rank of the energy content of each mode, indicating the prevalence of the corresponding flow structures in the field. The total number of extracted POD modes, $m$, is equal to the number of available snapshots of flowfield in the data here.

For a sampling plan with $p$ design parameters, the number of design settings $q$ can be determined by design of experiment (DoE) [42]. The vector form of the parametric values at design setting $i$ is denoted as $d_i \in R^p$. The spatiotemporal database of the flowfield $f(x,t;d_i)$ at design setting $i$ can be represented by a matrix $X_i \in R^{n \times m}$, where $m$ is the number of snapshots and $n$ is the total number of computational cells. The latter is usually much larger than the former, a situation particularly in large-scale numerical simulations of flowfields. The matrix $X_i$ can be uniquely factorized in the following form through singular value decomposition (SVD):

$$X_i = U_i \Lambda_i V_i^T. \qquad (2)$$

Here, $U_i$ is an $n \times n$ orthonormal matrix spanning the column space of $X_i$, $\Lambda_i$ an $n \times m$ diagonal matrix of singular values, and $V_i$ an $m \times m$ orthonormal matrix spanning the row space of $X_i$. $U_i$ and $V_i$ can be computed from the eigenvalues and eigenvectors of $X_i X_i^T$ and $X_i^T X_i$, respectively. Let $C_i$ be the inner product of the data matrix $X_i^T X_i$, also known as the *Gram matrix*.

The eigenvectors of $C_i$ make up the columns of $V_i$, i.e., $C_i = V_i L_i V_i^T$, where $L_i = \Lambda_i^T \Lambda_i \in R^{m \times m}$ is the diagonal matrix of eigenvalues of $C_i$.

All POD modes and time-varying coefficients of $X_i$ can then be written as:

$$\Phi_i = X_i V_i, \tag{3}$$

$$B_i = V_i, \tag{4}$$

where $\Phi_i = \{\phi_i^k(x), k = 1, 2, \cdots, m\} \in R^{n \times m}$, and $B_i = \{\beta_i^k(t), k = 1, 2, \cdots, m\} \in R^{m \times m}$. The reconstructed flowfield can be expressed as

$$X_i = \Phi_i B_i^T. \tag{5}$$

Model reduction is possible when the number of POD modes selected for reconstruction is truncated at a lower rank of $r$ ($r < m$). The POD technique is frequently used to develop projection-based surrogate models in computational fluid dynamics.

As mentioned in the Introduction, CPOD- and KSPOD surrogate models have been proposed for problems with varying flow conditions and geometries [17, 38]. CPOD concatenates all training information into a large matrix, $\mathbb{X} = (X_i, i = 1, 2, \cdots, q) \in R^{n \times mq}$, which is then used to build the CPOD covariance matrix and CPOD modes. To do this, a physics-guided common grid system is designed for the projection of the original database. A set of CPOD modes are obtained following SVD, and the corresponding time-varying coefficients at a new design setting are determined by the kriging procedure [17, 38]. On the other hand, the dimension of these CPOD modes ($mq$) is significantly larger than that of POD modes ($m$) at each training setting, producing an averaging effect on the flowfield. Consequently, the prediction accuracy of instantaneous flow information is compromised.

To overcome the limitations of CPOD, the KSPOD technique constructs spatial functions using the kriging-weighted average of POD modes at every sampling point, retaining dominant structures across all design points [39]. The method considerably improves prediction of instantaneous flowfield. A major difficulty, however, arises when the POD modes at different sampling points deviate in phase, as manifested by the element signs in the POD mode matrix. To this end, we propose a CKSPOD emulation to circumvent the issue of phase shift and provide faithful prediction of spatiotemporal flow dynamics in the present study. We now discuss the details of the proposed methodology.

In CKSPOD, a common Gram matrix $\mathbb{C}$, which contains data information of all training cases, is constructed as the Hadamard product of the Gram matrices of these cases. The Hadamard operator, denoted as $\circ$, is the element-to-element multiplication of two matrices of similar dimension, as follows.

$$\begin{bmatrix} a_{11} & a_{12} \\ a_{21} & a_{22} \end{bmatrix} \circ \begin{bmatrix} b_{11} & b_{12} \\ b_{21} & b_{22} \end{bmatrix} = \begin{bmatrix} a_{11}b_{11} & a_{12}b_{12} \\ a_{21}b_{21} & a_{22}b_{22} \end{bmatrix} \tag{6}$$

Accordingly, the common Gram matrix can be written as

$$\mathbb{C} = \boldsymbol{C}_1 \circ \boldsymbol{C}_2 \circ \ldots \circ \boldsymbol{C}_q = (\boldsymbol{V}_1 \boldsymbol{L}_1 \boldsymbol{V}_1^T) \circ \ldots \circ (\boldsymbol{V}_q \boldsymbol{L}_q \boldsymbol{V}_q^T) = \mathbb{V} \mathbb{L} \mathbb{V}^T, \tag{7}$$

where $\mathbb{V}$ is the column matrix of eigenvectors of $\mathbb{C}$ and $\mathbb{L}$ the corresponding diagonal matrix. The Gram matrix $\boldsymbol{C}_i$, the spatially averaged temporal correlation matrix, contains the temporal dynamics of the corresponding sampling point $i$, and the common Gram matrix synthesizes these dynamics through element-wise multiplication using the Hadamard product. This operation can adjust the phase-shift issue encountered in the previously proposed KSPOD model. Note that an inherent assumption applied in Eq. (7) is that the number of snapshots (*m*) collected for each design setting is identical. This ensures that all Gram matrices are of the same dimension, and enables the Hadamard product. For problems with different geometries and computational grids, an additional procedure is needed to project the original simulation data into a common grid space [17].

If we also define $\Pi$ as the notation of the Hadamard product, $\mathbb{C}$ can be organized as:

$$\mathbb{C} = \Pi_{j=1}^{q} \boldsymbol{C}_j = \boldsymbol{C}_i \circ \Pi_{\substack{j=1 \\ j \neq i}}^{q} \boldsymbol{C}_j = \boldsymbol{V}_i \boldsymbol{L}_i \boldsymbol{V}_i^T \circ \Pi_{\substack{j=1 \\ j \neq i}}^{q} \boldsymbol{C}_j. \tag{8}$$

Directly using the POD modes from the original Gram matrix of each training case, as in KSPOD, the CKSPOD technique implements the common Gram matrix in Eq. (8) to deduce spatial functions and time-varying coefficients for the training cases. The transferred spatial functions of the design setting $i$, denoted as $\boldsymbol{\Phi}_i'$, are written as

$$\boldsymbol{\Phi}_i' = \boldsymbol{X}_i \mathbb{V} = \boldsymbol{X}_i \mathbb{C} \mathbb{V} \mathbb{L}^{-1}. \tag{9}$$

Substituting $\mathbb{C}$ using Eq. (8), $\boldsymbol{\Phi}_i'$ can be related to the original POD modes $\boldsymbol{\Phi}_i$ in the following form

$$\boldsymbol{\Phi}_i' = \boldsymbol{X}_i \boldsymbol{V}_i \left\{ \left[ \boldsymbol{L}_i \boldsymbol{V}_i^T \circ \left( \Pi_{\substack{j=1 \\ j \neq i}}^{q} \boldsymbol{C}_j \right) \right] (\mathbb{V} \mathbb{L}^{-1}) \right\} = \boldsymbol{X}_i \boldsymbol{V}_i \boldsymbol{\mathcal{T}} = \boldsymbol{\Phi}_i \boldsymbol{\mathcal{T}} \tag{10}$$

Here the matrix $\boldsymbol{\mathcal{T}}$ is defined as

$$\boldsymbol{\mathcal{T}} = \left[ \boldsymbol{L}_i \boldsymbol{V}_i^T \circ \left( \Pi_{\substack{j=1 \\ j \neq i}}^{p} \boldsymbol{C}_j \right) \right] \mathbb{V} \mathbb{L}^{-1}. \tag{11}$$

We call $\boldsymbol{\mathcal{T}}$ the CKSPOD transfer matrix, which converts the original POD modes to the spatial functions at each design setting. To utilize these spatial functions properly with kriging-based weights for the new design setting, a normalization process is performed.

$$\widetilde{\boldsymbol{\Phi}}_i = \frac{\boldsymbol{\Phi}_i'}{\|\boldsymbol{\Phi}_i'\|_2} \tag{12}$$

Accordingly, the transferred matrix of time-varying coefficients is expressed as

$$\widetilde{\boldsymbol{B}}_i = \mathbb{V}^T \|\boldsymbol{\Phi}'_i\|_2. \tag{13}$$

Once the transferred matrices of new spatial functions and time-varying coefficients are deduced, kriging is implemented to develop the basis functions and coefficients for a new design setting. It should be noted that the transferred spatial functions $\widetilde{\boldsymbol{\Phi}}_i$ maintain the orthogonality after the conversion by the CKSPOD transfer matrix, and can thus be considered as new POD modes.

### 2.2. The kriging model

Kriging, also known as Gaussian process (GP) regression, models the responses over the input space as a sample drawn from a GP. For a set of design settings $\{\boldsymbol{d}_i \in R^p\}_{i=1}^q$, the observed functions of interest are weighting parameters of spatial basis modes and corresponding mode coefficients. Based on the training dataset of the $\boldsymbol{d}_i - \boldsymbol{y}_i$ pair, with a new input $\boldsymbol{d}_{new}$, kriging predicts the corresponding response $\boldsymbol{y}_{new}$. The mathematical formula of kriging in terms of prediction $\boldsymbol{y}_{new}$ is given by

$$\boldsymbol{y}_{new} = \mathbb{E}[y(\boldsymbol{d}_{new})|\boldsymbol{y}] = \hat{\boldsymbol{\mu}} + \boldsymbol{r}_{new}^T \boldsymbol{R}^{-1}(\boldsymbol{y} - \boldsymbol{1}_n \hat{\mu}) \tag{14}$$

where $\hat{\boldsymbol{\mu}} = \boldsymbol{1}_q^T \boldsymbol{R}^{-1} \boldsymbol{y} / \boldsymbol{1}_q^T \boldsymbol{R}^{-1} \boldsymbol{1}_q$ is the maximum likelihood estimate of $\boldsymbol{\mu}$, $\boldsymbol{1}_q$ is a $q$–vector of 1's, and $\boldsymbol{R}$ is a $q \times q$ matrix of re-parameterized squared-exponential correlation function whose (i, j)-th entry is $r(\boldsymbol{d}_i, \boldsymbol{d}_j) = \exp\left\{-\sum_{k=1}^p \theta_k (d_{kj} - d_{ki})^2\right\}$ with $\theta_k = -4 \log d_k$. $\boldsymbol{r}_{new}$ is a $p$-vector whose $i$-th entry is $r(\boldsymbol{d}_{new}, \boldsymbol{d}_i)$. This allows for a more numerically stable optimization of maximum likelihood estimators [17].

Replacing $\boldsymbol{y}$ in Eq. (14) with the column vectors of $\widetilde{\boldsymbol{B}}_i$ ($\tilde{\beta}_i^k, k = 1, \cdots, m$), the predicted time-varying coefficients at unobserved design setting $\widehat{\boldsymbol{B}}_{new}$ ($\hat{\beta}_{new}^k, k = 1, \cdots, m$) can be obtained. The spatial basis functions are calculated in a slightly different way. Kriging is used to predict the weight of each spatial function at observed points on the spatial function at the new design setting.

To this end, the observations $y$ are now taken to be the orthonormal vector $e_i$, where $e_i$ is a $q$-vector with 1 in its $i$-th element and 0 elsewhere. Intuitively, this quantifies the fact that the spatial mode information extracted in the $i$-th design setting corresponds to only that setting and not the other $q-1$ settings. With this in mind, the resulting predictor in Eq. (14) can be viewed as the predicted weight for that particular spatial mode at the new design setting $d_{new}$, denoted as $\widehat{w}_{new,i}$. This procedure is repeated for each of the $p$ unit vectors $(e_i)_{i=1}^{q}$, from which the $q$ weighting parameters $(\widehat{w}_{new,i})_{i=1}^{q}$ can be obtained. The weighting parameters are normalized to ensure their summation equal to unity.

The weighting parameters are subsequently used to predict the new spatial function modes through a weighted average of the extracted modes at the new design settings, expressed as,

$$\widehat{\phi}^k(d_{new}, x) = \sum_{i=1}^{p} \widehat{w}_{new,i} \widetilde{\phi}_i^k \tag{15}$$

With the time-varying coefficients and spatial functions obtained using Eqs. (14)-(15), the predicted spatiotemporal flowfield at a new design setting is

$$\widehat{X}(d_{new}, x, t) = \sum_{k=1}^{m} \widehat{\beta}^k(d_{new}, t) \widehat{\phi}^k(d_{new}, x) \tag{16}$$

Based on the mathematical formulation described above, Algorithm 1 outlines the steps of the CKSPOD approach. First, prepare the data matrix from numerical simulations. Second, construct the common Gram matrix based on the Gram matrix at each observed design setting using the Hadamard-based product. Third, build the transfer matrix and create new spatial functions and corresponding coefficients at all design settings. Fourth, introduce kriging to establish spatial functions and time-varying coefficients at a new design setting. Finally, construct the spatiotemporal flowfield at the new design setting.

Note that the KSPOD approach [39] directly uses the original POD modes ($\Phi_i$) and coefficient ($B_i$) at each design setting $i$ to perform the kriging process using Eq. (14). Correspondingly Equations (15) and (16) are updated using different spatial functions and time-varying coefficients at the new design setting. For comparison, the results of the KSPOD approach are also presented in the following section.

**Algorithm 1. Common kernel-smoothed POD (CKSPOD) surrogate model**

| | | |
|---|---|---|
| **DATA:** | | For each design point in $\{d_i\}_{i=1}^{p}$, prepare spatiotemporally evolving flowfield data matrix $X_i$ from numerical simulations. |
| **TRAINING:** | Step 1: | Construct the common Gram matrix $\mathbb{C}$ based on Gram matrices of all observed design settings using Hadamard product (Eq.(8)). |
| | Step 2: | Build transfer matrix $\mathcal{T}$ using Eq. (11) and create spatial functions and coefficients at all design settings (Eqs. (12)-(13)). |
| | Step 3: | For each time step $t_q$ and each mode $k$, fit the kriging model with responses $\{\tilde{\beta}^k(d_1,t_q),...,\tilde{\beta}^k(d_p,t_q)\}$ at observed design settings of $\{d_1,...,d_p\}$, and the predictive coefficients for a new design setting $d_{new}$ is $\hat{\beta}^k(d_{new},t_q)$ (Eq. (14)). |
| | Step 4: | Perform the kriging model with inputs $\{d_1,...,d_p\}$ and the predictive function for $d_{new}$ is the weight $\hat{w}_i(d_{new})$ associated with each design setting $i$. The predicted spatial function of mode |

| | $k$, $\widehat{\boldsymbol{\phi}}^k(\boldsymbol{c}_{new}, \boldsymbol{x})$ is the weighted sum of corresponding spatial functions at observed design settings (Eq. (15)). |
|---|---|
| **PREDICTION:** | Use Eq. (16) to construct the spatiotemporal flowfield data at the new design setting. |

## 3. Uncertainty quantification

For surrogate modeling, it is also crucial to quantify the uncertainty of the prediction to assess the model accuracy. Here, we assume that the database created by high-fidelity LES simulations is reliable and accurate, and the uncertainty of statistical prediction primarily results from the kriging process. Our previous work in CPOD has shown that invoking the conditional distribution of the multivariate normal distribution, the kriging-predicted time-varying coefficients at a new design setting follow the Gaussian distribution,

$$\boldsymbol{\beta}(\boldsymbol{d}_{new})|\{\widetilde{\boldsymbol{\beta}}(\boldsymbol{d}_i)\}_{i=1}^q \sim N(\widehat{\boldsymbol{\beta}}, \widehat{\boldsymbol{\Sigma}}) \tag{17}$$

Here the minimum mean square error (MMSE) predictor $\widehat{\boldsymbol{\beta}}$ for $\boldsymbol{\beta}(\boldsymbol{d}_{new})|\{\widetilde{\boldsymbol{\beta}}(\boldsymbol{d}_i)\}_{i=1}^q$ and its corresponding variance are given by,

$$\widehat{\boldsymbol{\beta}}(\boldsymbol{d}_{new}) = \boldsymbol{\mu} + \left((\boldsymbol{r}_{new}^T \boldsymbol{R}^{-1}) \otimes \boldsymbol{I}_m\right)(\widetilde{\boldsymbol{\beta}} - \boldsymbol{1}_p \otimes \boldsymbol{\mu}) \tag{18}$$

$$\widehat{\boldsymbol{\Sigma}} = \mathbb{V}\{\boldsymbol{\beta}(\boldsymbol{d}_{new})|\{\boldsymbol{\beta}(\boldsymbol{d}_i)\}_{i=1}^q\} = (1 - \boldsymbol{r}_{new}^T \boldsymbol{R}^{-1} \boldsymbol{r}_{new}) \boldsymbol{T} \tag{19}$$

where $\boldsymbol{I}_m$, $\boldsymbol{1}_n$, and $\boldsymbol{T}$ are $m \times m$ identity matrix, 1-vector of $q$ elements, and the $m \times m$ covariance matrix, respectively.

Similarly, the variance associated with the weights of the predicted spatial functions during kriging at a new design setting can also be represented by Eq. (21), except with $T$ as an identity matrix. The UQ of the final prediction using Eq. (16) can be calculated through the propagation of the uncertainties of the weights for the time-varying coefficients and the spatial functions. The spatiotemporal variance is expressed as

$$\mathbb{V}\{X(x,t; d_{new})\}|\{X(x,t; d_i)\}_{i=1}^{q}$$
$$= \sum_{k=1}^{m} \mathbb{V}\left\{\widetilde{\beta}(d_{new})|\{\widetilde{\beta}(d_i)\}_{i=1}^{q}\right\} \sum_{i=1}^{q} \mathbb{V}\{\widehat{w}_i(d_{new})|\{w(d_i)\}_{i=1}^{q}\}\{\widetilde{\phi}_i^k(\mathbf{x})\}^2 \quad (20)$$

The spatiotemporal variance of the reconstructed flowfield using KSPOD follows a similar format, but the transferred spatial functions are replaced by the original POD modes of the observed design settings ($\{\phi_i^k(x)\}_{i=1}^{q}$).

To define and illustrate the UQ of the proposed models the derived quantity turbulent kinetic energy (TKE) is implemented, defined as

$$\kappa(\mathbf{x}, t) = \frac{1}{2} \sum_{i \in (x,y,z)} \{v_i(x,t) - \bar{v}_i(x,t)\}^2, \quad (21)$$

where $v_x(x,t)$, $v_y(x,t)$, and $v_z(x,t)$ correspond to velocity components in the $x$, y, and $z$ directions, respectively. $\kappa(\mathbf{x}, t)$ is important, because it is able to measure the energy within turbulent eddies and vortices. Therefore, the MMSE predictor and pointwise CI for $\kappa(x,t)$ (see Theorem 1 in Ref. [17]) at a new design point $d_{new}$, can be computed by combining Eq. (21) and the variable of each predicted model. This type of computation of the distribution function has been extensively studied [43, 44], and these methods are utilized for computing the pointwise CI of $\kappa(\mathbf{x}, t)$ in Section III. With this in hand, the prediction and UQ of TKE from the CKSPOD and KSPOD are compared with the simulated TKE at the validation design point.

## 4. Model validation: Swirl injectors

The proposed CKSPOD emulation methodology is applied to the design of a swirl injector. Swirl injectors have been widely used to achieve efficient mixing and combustion in propulsion and power-generation systems [1, 16]. The performance of the CKSPOD technique is examined in detail in this section.

### 4.1. Swirl injector configuration

Figure 1 shows sectional views of the simplex swirl injector of concern [16-18], identical to the one used in our previous KSPOD work [39]. Liquid oxygen (LOX) is tangentially introduced into the injector through orifices. In this study, the orifices are replaced by a slit on the injector wall; the slit width ($\delta$) is carefully chosen to ensure a mass flow rate identical to that of the discrete orifices. A liquid film forms along the injector wall [45-48]. A hollow gaseous core is formed in the center region due to the conservation of angular momentum. The liquid film exits the injector as a thin conical sheet and subsequently undergoes rapid atomization. The detailed flow dynamics of this type of simplex swirl injector under supercritical conditions has previously been thoroughly studied using LES [46, 47].

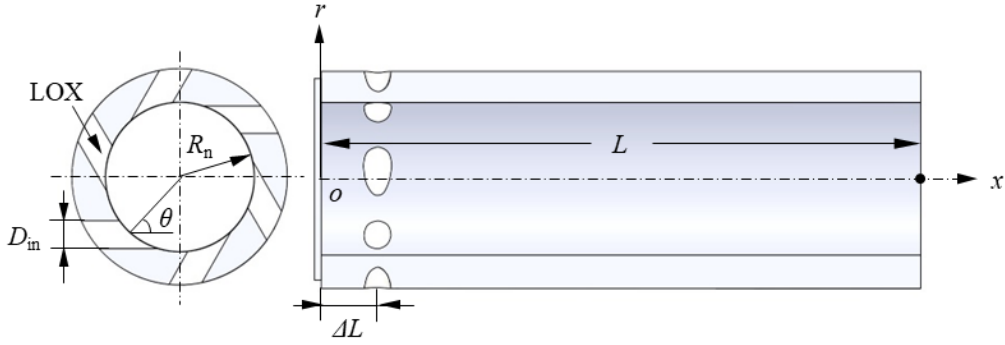

**Fig. 1. Schematic of swirl injector**

The selection of injector design variables requires careful assessment of their influences on flow dynamics. The major design attributes include the injector length, $L$, radius, $R_n$, tangential inlet width, $\delta$, tangential injection angle, $\theta$, and distance between the inlet and headend, $\Delta L$. A sensitivity analysis described in previous studies [17, 38] has identified $\delta$, $\theta$, and $\Delta L$ as the most significant parameters. They are thus selected as the design parameters in the current study. The baseline geometry and operating conditions of the injector, including the LOX inlet temperature $T_{in}$, ambient temperature $T_\infty$, ambient pressure $p_\infty$, and mass flow rate $\dot{m}$, are tabulated in Table 1.

The range for each design variable is listed in Table 2. The distance between the inlet and headend, $\Delta L$, is typically 1.5-2.0 times the diameter of the injection orifice [45]; the location depends on a tradeoff to avoid (1) excessive viscous losses when the injection slit is too close to the headend, and (2) low frequency oscillations due to the presence of a recirculation zone if the inlet is too far from the headend. The design ranges for the injector width $\delta$ and angle $\theta$ are selected based on the desired values of the film spreading angle (50-62°) and thickness (0.66-1.50 mm) at the injector exit.

With the selected design parameters and their ranges, the SLHD method is implemented to generate the design settings with the $n=10p$ rule-of-thumb for sample size [49]. A total number of 30 design settings are employed in the present study, as listed in Table 3 [32]. The three design parameters are significantly influenced by the inlet velocity, $u_{in}$, ranging from 5.71 to 40.43 m/s. The 30 training cases are thus classified into four groups in terms of $u_{in}$ (m/s) as follows: Cluster A with $u_{in} < 10$; Cluster B with $10 \leq u_{in} < 18$; Cluster C with $18 \leq u_{in} < 25$; and Cluster D with $u_{in} > 25$.

In SLHD, the space-filling property of the design points in each slice is optimal. The overall design matrix contains five slices, and each slice includes six design settings. Fig. 2 shows two-dimensional projections of the design settings categorized by different slices. LES-based high-fidelity simulations are performed at all design settings.

Table 1. Baseline injector geometry and operating conditions

| $R$ (mm) | $D_{in}$ (mm) | $L$ (mm) | $\dot{m}$ (kg/s) | $T_{in}$ (K) | $T_{\infty}$ (K) | $p_{\infty}$ (MPa) |
|---|---|---|---|---|---|---|
| 4.50 | 1.70 | 25 | 0.17 | 120 | 300 | 10 |

Table 2. Design space

| Design Variable | $\theta$ (deg) | $\delta$ (mm) | $\Delta L$ (mm) |
|---|---|---|---|
| Design Range | 35.0-62.2 | 0.27-1.53 | 0.85-3.40 |

Table 3. Design matrix and associated inlet velocity information

| Case | $\delta$ (mm) | $\theta$ (deg) | $\Delta L$ (mm) | $u_{in}$ (m/s) | $u_r$ (m/s) | $u_\theta$ (m/s) | $K$ | Cluster |
|---|---|---|---|---|---|---|---|---|
| 1 | 0.28 | 57.92 | 1.59 | 40.43 | 21.47 | 34.26 | 7.44 | D |
| 2 | 0.63 | 40.81 | 1.93 | 12.35 | 9.35 | 8.07 | 1.64 | B |
| 3 | 0.82 | 52.39 | 0.96 | 11.79 | 7.20 | 9.34 | 1.98 | B |
| 4 | 1.10 | 32.76 | 2.57 | 6.42 | 5.40 | 3.47 | 0.69 | A |
| 5 | 1.12 | 51.88 | 3.21 | 8.58 | 5.30 | 6.75 | 1.43 | A |
| 6 | 1.52 | 46.85 | 2.23 | 5.71 | 3.90 | 4.16 | 0.86 | A |
| 7 | 0.38 | 37.29 | 1.64 | 19.53 | 15.54 | 11.83 | 2.37 | C |
| 8 | 0.51 | 52.89 | 2.15 | 19.35 | 11.67 | 15.43 | 3.27 | C |
| 9 | 0.78 | 43.33 | 3.12 | 10.43 | 7.58 | 7.15 | 1.46 | B |
| 10 | 1.03 | 33.76 | 0.87 | 6.89 | 5.73 | 3.83 | 0.76 | A |
| 11 | 1.26 | 49.37 | 1.72 | 7.19 | 4.68 | 5.46 | 1.14 | A |
| 12 | 1.39 | 60.44 | 2.61 | 8.63 | 4.26 | 7.51 | 1.65 | A |

| | | | | | | | |
|---|---|---|---|---|---|---|---|
| 13 | 0.47 | 54.40 | 2.74 | 21.87 | 12.73 | 17.78 | 3.80 | C |
| 14 | 0.68 | 38.80 | 2.53 | 11.25 | 8.77 | 7.05 | 1.42 | B |
| 15 | 0.74 | 48.36 | 1.89 | 12.06 | 8.02 | 9.02 | 1.88 | B |
| 16 | 0.93 | 33.26 | 1.47 | 7.63 | 6.38 | 4.18 | 0.83 | A |
| 17 | 1.22 | 42.82 | 0.91 | 6.60 | 4.84 | 4.49 | 0.92 | A |
| 18 | 1.35 | 57.42 | 3.17 | 8.15 | 4.39 | 6.87 | 1.49 | A |
| 19 | 0.32 | 58.43 | 2.27 | 35.58 | 18.63 | 30.31 | 6.60 | D |
| 20 | 0.59 | 34.77 | 1.13 | 12.19 | 10.01 | 6.95 | 1.38 | B |
| 21 | 0.84 | 49.87 | 2.83 | 10.89 | 7.02 | 8.32 | 1.74 | B |
| 22 | 0.99 | 44.33 | 1.76 | 8.35 | 5.97 | 5.84 | 1.20 | A |
| 23 | 1.20 | 37.79 | 3.08 | 6.24 | 4.93 | 3.82 | 0.77 | A |
| 24 | 1.45 | 55.41 | 1.55 | 7.17 | 4.07 | 5.90 | 1.27 | A |
| 25 | 0.40 | 36.28 | 2.32 | 18.27 | 14.73 | 10.81 | 2.16 | C |
| 26 | 0.49 | 51.38 | 1.42 | 19.51 | 12.18 | 15.24 | 3.21 | C |
| 27 | 0.72 | 53.39 | 3.29 | 13.84 | 8.25 | 11.11 | 2.36 | B |
| 28 | 0.95 | 40.31 | 1.17 | 8.18 | 6.24 | 5.29 | 1.07 | A |
| 29 | 1.24 | 59.43 | 1.98 | 9.36 | 4.76 | 8.06 | 1.76 | A |
| 30 | 1.37 | 43.83 | 2.78 | 5.99 | 4.32 | 4.15 | 0.85 | A |

Note: Cases 1-6 are on Slice 1, Cases 7-12 on Slice 2, Cases 13-18 on Slice 3, Cases 19-24 on Slice 4, and Cases 25-30 on Slice 5, corresponding to the symbols on Fig. **Error! Reference source not found.**.

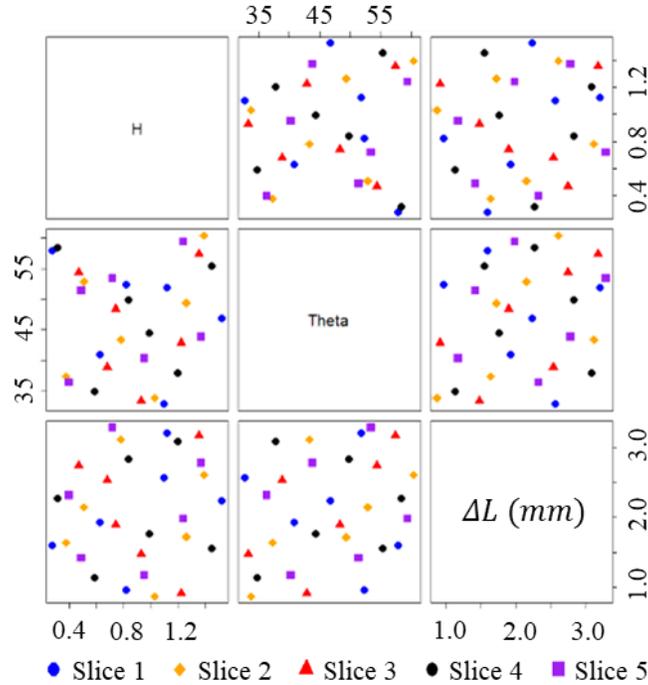

**Fig. 2. Two-dimensional projection of design points obtained by Sliced Latin Hypercube Design (SLHD) methodology in the design space.**

### 4.2. High-fidelity simulation

The theoretical formulation for high-fidelity simulations is described in detail in [50]. It allows for detailed modeling of mixing and combustion over the entire range of fluid thermodynamic states of concern [51]. Turbulence closure is achieved by means of an LES technique. Thermodynamic properties are evaluated according to fundamental thermodynamics theories and a modified Soave-Redlich-Kwong (SRK) equation of state (EOS). Transport properties, including thermal conductivity and viscosity, are determined using extended corresponding-state principles. Mass diffusivity is estimated based on the Takahashi method, calibrated for high-pressure conditions.

The numerical framework is based on a preconditioning scheme with a unified treatment of general-fluid thermodynamics [50, 52]. It utilizes a density-based, finite-volume methodology, along with a dual-time-step integration technique. Temporal discretization is accomplished by a second-order backward difference; and the inner-loop pseudo-time term is integrated via a four-step Runge-Kutta scheme. A fourth-order central difference scheme in generalized coordinates is used to obtain spatial discretization. Fourth-order matrix dissipation is applied to secure numerical stability with minimum contamination of the solution. Finally, a multi-block domain decomposition technique associated with the message passing interface technique for parallel computing is applied to optimize computation performance.

### 4.3. POD modes and CKSPOD spatial functions

With spatiotemporal simulation data assigned for all 30 design settings, a database is established to train the emulator. To justify the applicability of CKSPOD, the original POD modes are matched to the transferred spatial functions deduced from the common covariance matrix, aliased as the CKSPOD modes for convenience. Figure 3 shows the first four modes of POD and CKSPOD for the pressure field of Case 16. All modes are normalized to facilitate comparison. The POD and CKSPOD modes exhibit similar features, which suggests that most significant coherent flow structures in the original design setting are retained after the CKSPOD procedure. The energy percentage of each POD mode, however, is consistently larger than its CKSPOD counterpart. Mathematically, the energy percentage is represented by the ratio of the diagonal element (eigenvalue) to the trace of the diagonal matrix $\Lambda$. The energy carried by the fist POD mode is around 2.5 times of the corresponding CKSPOD mode; this can be explained by the introduction of the CKSPOD transfer matrix defined in Eq. (11). This transfer matrix contains the dominant dynamics embedded in the other 29 design settings, and therefore reduces the energy percentage of the original POD mode.

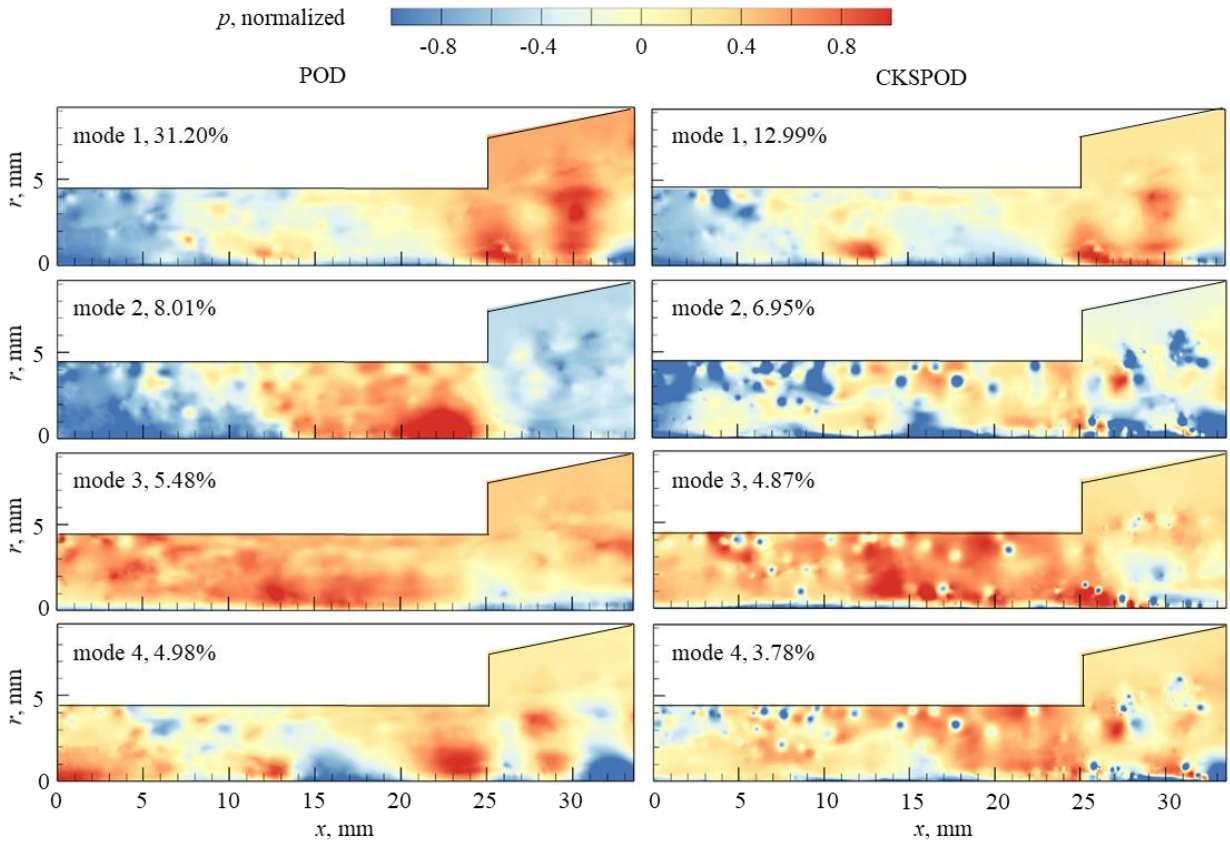

**Fig. 3. Pressure POD Modes 1-4 for Case 16 from Cluster A and corresponding CKSPOD modes.**

To further demonstrate the similarity, Figure

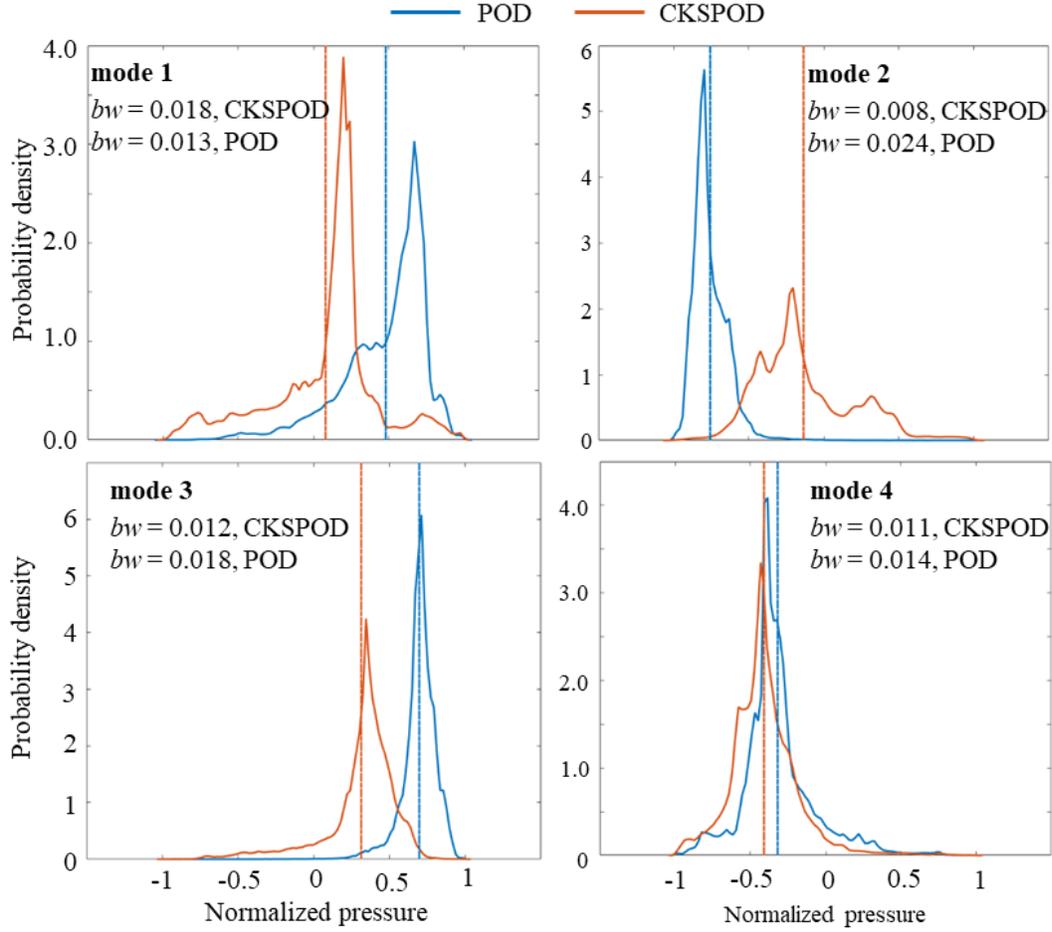

Fig. 4 shows probability density distributions of the first four pressure modes of POD and CKSPOD in Case 16 based on the kernel smoothing function. A kernel distribution is a nonparametric representation of the probability density function, $f_h(x)$, of a random variable, written as

$$f_h(x) = \frac{1}{nh}\sum_{i=1}^{n}\mathcal{K}\left(\frac{x-x_i}{bw}\right) \qquad (22)$$

where $n$ the sample size, $\mathcal{K}$ the kernel density smoothing function, and $bw$ the bandwidth acting as a smoothing parameter. Here, the bandwidth of the kernel smoothing function is optimally selected for estimating densities for the normal distribution [53] to produce a reasonably smooth curve. Close similarity is observed between the probability density distributions of the POD and CKSPOD modes. Both Modes 1 and 3 show negative skew with a tail on the left. The distributions of the fourth POD and CKSPOD modes are almost overlapped, with the latter slightly shifted to the left. The transfer matrix changes the mean value of the distribution and rescales the magnitude the modes. Such behavior is critical so POD modes for all design settings can be automatically adjusted, unlike the situation with the KSPOD approach [39], in which all POD modes must be manually checked to avoid the occurrence of the kriging procedure.

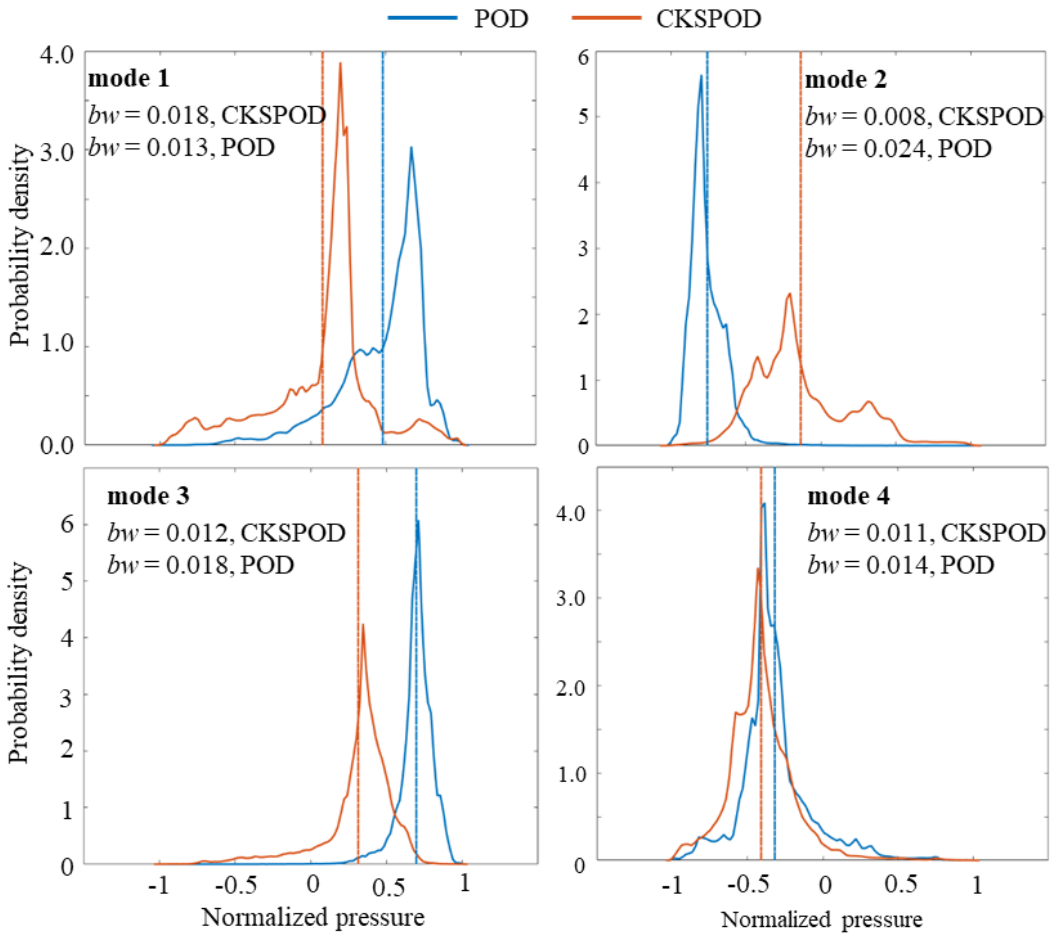

**Fig. 4. Probability densities of POD Modes 1-4 for Case 16 from Cluster A and corresponding modes. Vertical lines represent mean values and *bw* represents bandwidth of kernel smoothing function.**

Figure 5 shows the accumulated energy percentage of the POD modes for the pressure field for 8 randomly selected cases from the 30 design settings. Also included are the results for the CKSPOD modes. In the CKSPOD approach, all of the 30 cases are treated using the identical eigen-decomposition process with the same energy accumulation through the CKSPOD transfer matrix. The phase-difference issue encountered in the KSPOD method is thus remedied. On the other hand, the energy accumulation is much slower for the CKSPOD modes, due to the data smoothing effect. It takes the first 173 modes to obtain 90% of the total energy for the CKSPOD method, compared to the first 23-99 modes for the POD method. More modes are required for the CKSPOD approach to capture the same amount of energy in the establishment of the emulator. In the present study, the first 522 modes, which cover more than 99% percentage of the total energy, are included to the train the surrogate model and build the spatial functions and coefficients at the new design setting.

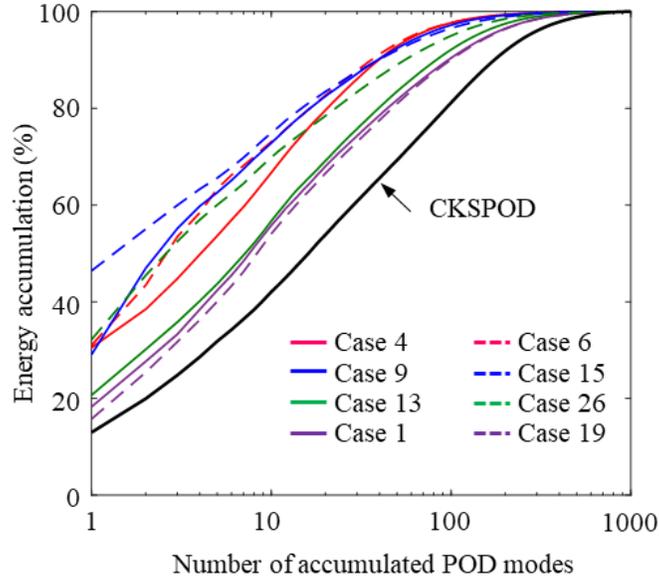

**Fig. 5. Accumulated energy percentage of POD modes for the pressure field**

### 4.4. Prediction by CKSPOD emulation

Eight new design settings within the design space are selected as the validation cases. The parameter settings are carefully determined to cover a wide range of inlet velocity $u_{in}$ from 5.71 to 40.43 m/s. The eight new design settings are categorized into four groups according to the inlet velocity range: A-low; B-intermediate low; C-intermediate high; D-high inlet velocity, as listed in Table 4. Each group consists of two validation cases to fully evaluate the performance of the CKSPOD emulation. High-fidelity simulations are also performed at these settings to validate the predicted results by the emulation.

**Table 4. Design parameters for eight test cases in four different groups**

| Case | $\delta$ (mm) | $\theta$ (deg) | $\Delta L$ (mm) | $u_{in}$ (m/s) | $u_r$ (m/s) | $u_\theta$ (m/s) |
|---|---|---|---|---|---|---|
| A1 | 1.26 | 44.11 | 0.94 | 6.55 | 4.70 | 4.56 |
| A2 | 1.20 | 41.97 | 0.90 | 6.65 | 4.94 | 4.44 |
| B1 | 0.70 | 40.73 | 2.71 | 11.12 | 8.43 | 7.26 |
| B2 | 0.71 | 52.59 | 3.24 | 13.79 | 8.38 | 10.95 |
| C1 | 0.42 | 37.73 | 2.41 | 17.91 | 14.16 | 10.96 |
| C2 | 0.49 | 57.12 | 2.88 | 22.33 | 12.12 | 18.75 |
| D1 | 0.27 | 50.39 | 1.40 | 34.37 | 21.91 | 26.48 |
| D2 | 0.33 | 60.76 | 2.32 | 36.32 | 17.74 | 31.70 |

Figure 6 shows the instantaneous density fields of the LES-based simulation and the CKSPOD-emulation for Case D2. The evolution of the liquid film (seen in the dense fluid) and its spreading downstream of the injection port agree well between the simulation and emulation.

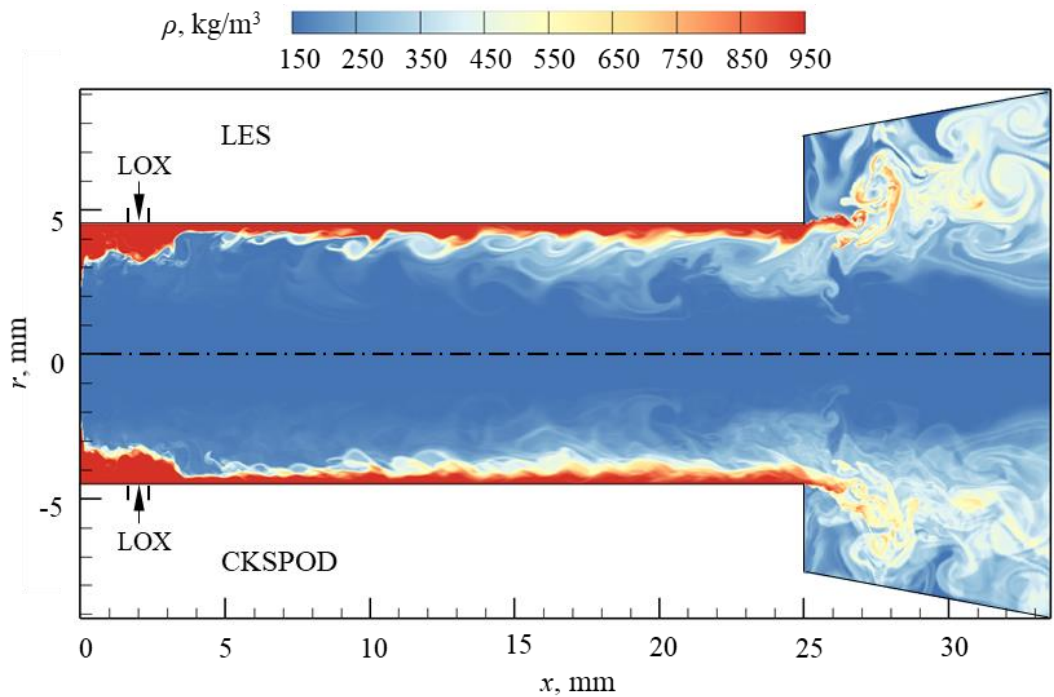

**Fig. 6. Comparison of density fields: LES-based simulation and CKSPOD emulation. Test Case D2 at t = 7.89 ms.**

4.4.1. Spatial distribution

The instantaneous spatial distributions are systematically examined to evaluate the emulator performance. Figures 7 and 8 show snapshots of the density field predicted by the LES, CPOD, KSPOD, and CKSPOD methods for Cases A2 and C2, respectively. In both cases, the CPOD method leads to blurred patterns in the stratified density layer, where the wavy structures are smoothed. Downstream of the injector, large-scale vortical motions calculated by the LES technique are significantly dissipated in the CPOD case. The results are consistent with the previous finding that the CPOD method can accurately capture major performance metrics, such as time-averaged liquid film thickness and spreading angle, but not detailed flow dynamics [10,11].

The KSPOD method accurately emulates the wavy structures of the liquid film; however, many small-scale motions within the injector and vortical structures downstream of the injector exit are not captured well, compared to the LES result. The CKSPOD method significantly improves the prediction and captures flow motions at all scales. The stringy ligaments and small vortices are faithfully emulated. In addition, the spreading of the liquid film and associated growth of vortical structures downstream of the injector are predicted with high fidelity.

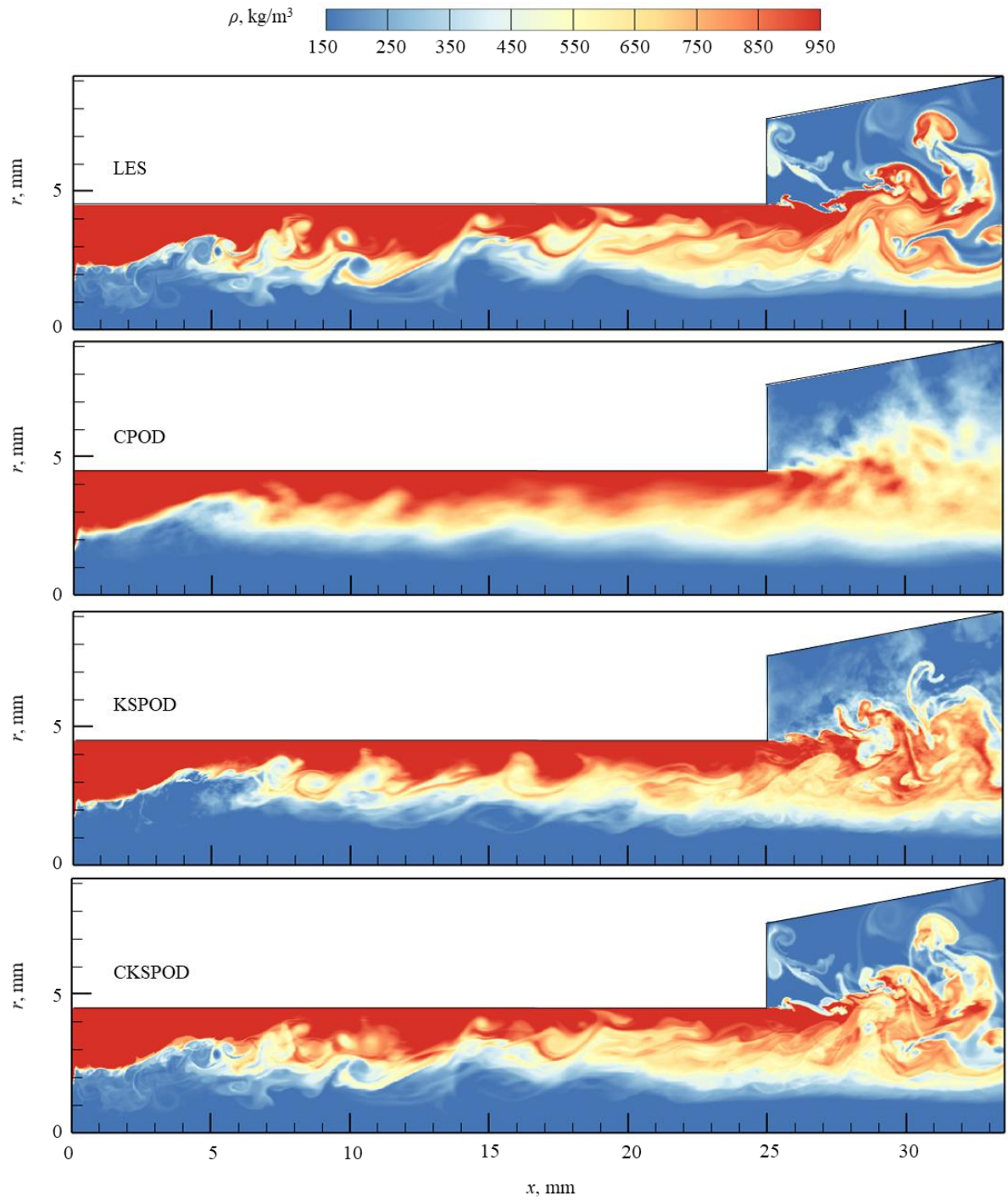

**Fig. 7. Comparison of density field: LES-based simulation and predictions by three different emulations. Test Case A2 at t = 1.01 ms.**

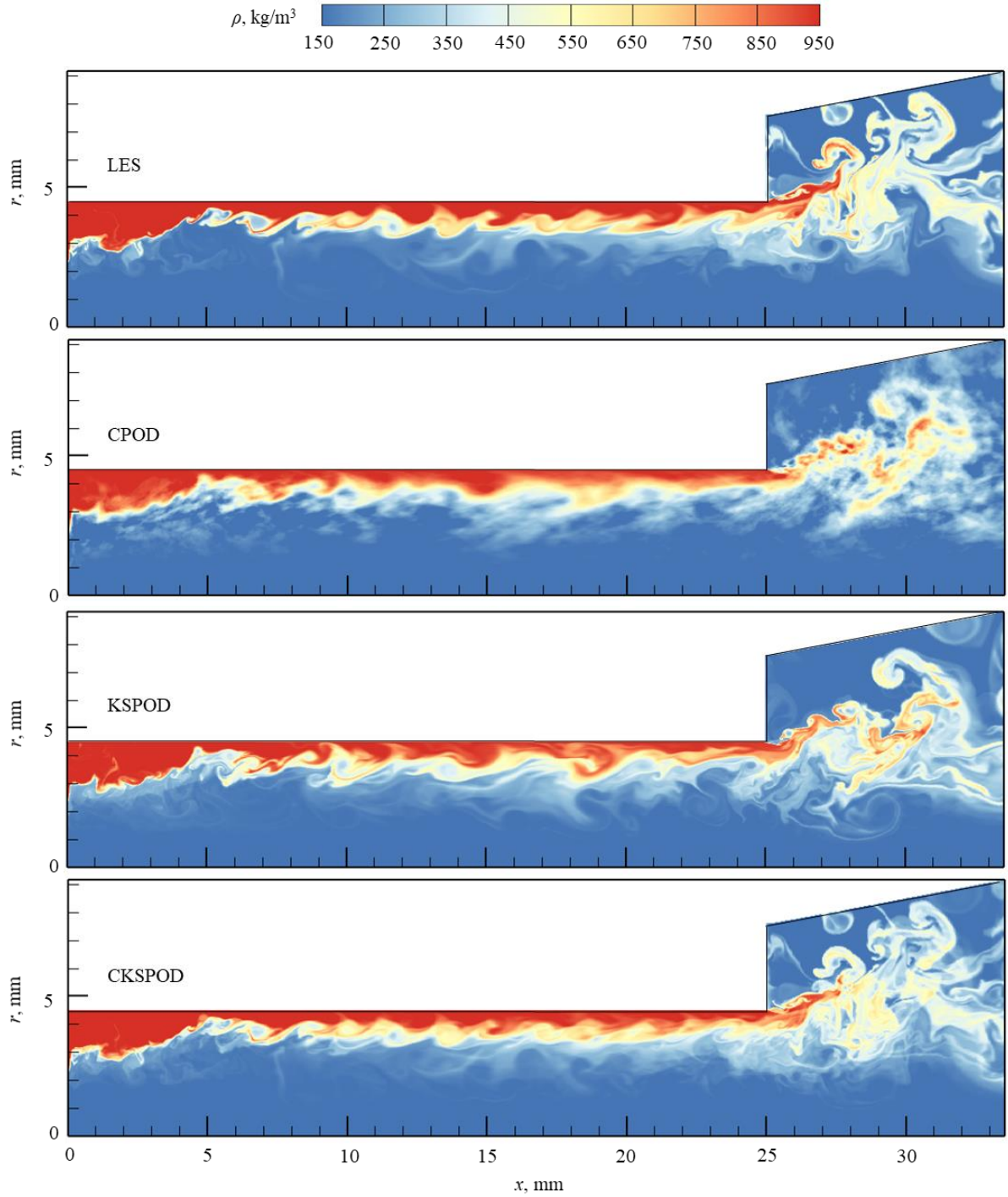

**Fig. 8. Comparison of density field: LES-based simulation and predictions by three different emulations. Test Case C2 at t = 0.11 ms.**

To further evaluate the performance and applicability of the CKSPOD method for prediction of the spatial structures of the flowfield, the density field in test cases B1 and B2 are presented in Figs. 9 and 10, respectively, for LES, KSPOD, and CKSPOD methods. Consistent trends of flow structure prediction in the CKSPOD method are observed here. The consistent improvement of the predictive capability of the CKSPOD method over the CPOD and KSPOD methods are clearly observed over a broad range of flow conditions.

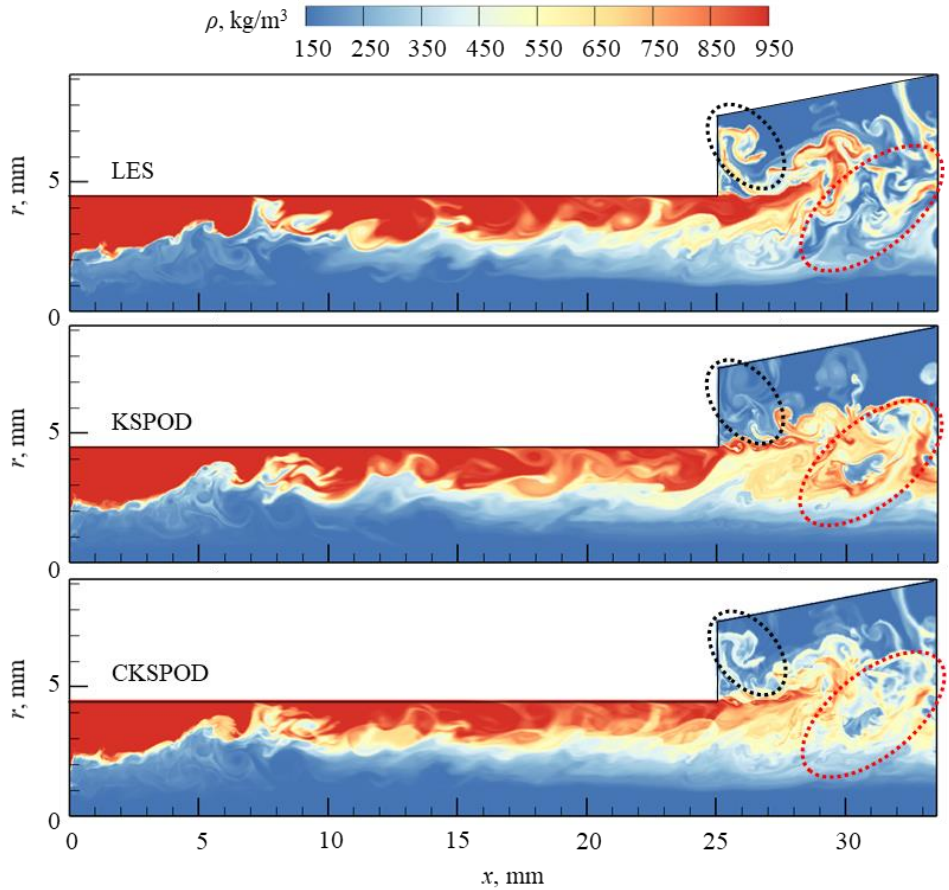

**Fig. 9. Comparison of density field: LES-based simulation and predictions by KSPOD- and CKSPOD emulations. Test Case B1 at t = 4.62 ms.**

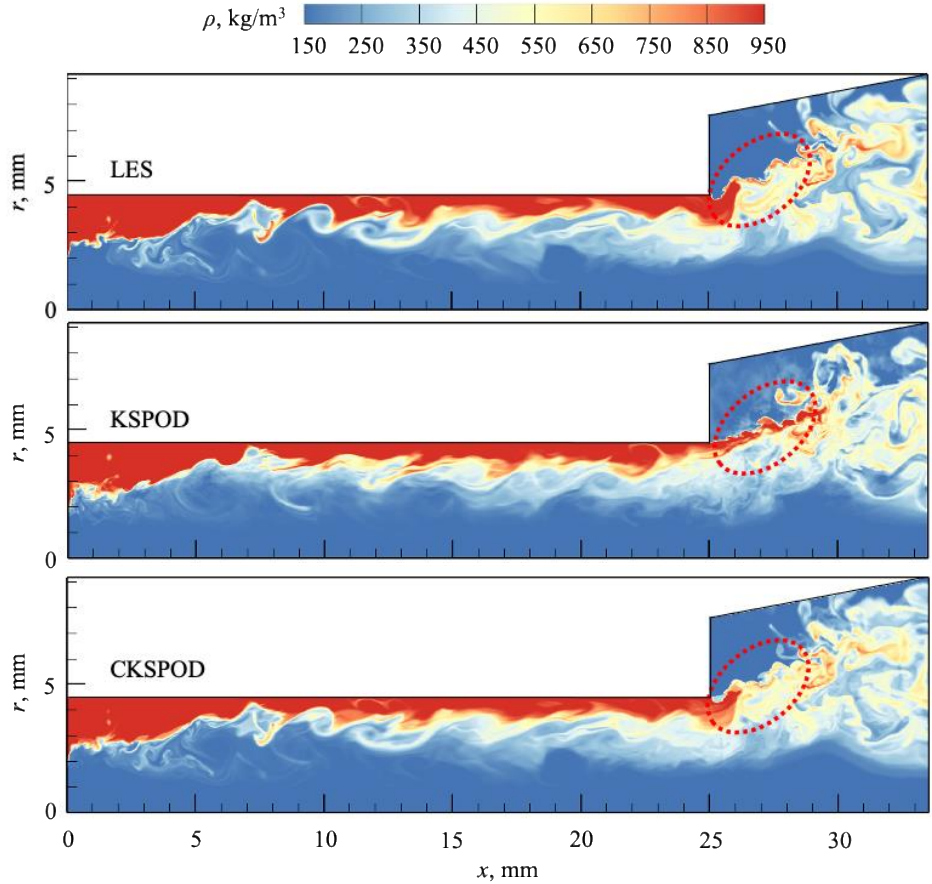

**Fig. 10. Comparison of density field: LES-based simulation and predictions by KSPOD and CKSPOD emulations. Test Case B2 at t = 2.13 ms.**

4.4.2. Temporal evolution

The performance of the CKSPOD method is further examined in terms of the temporal evolution of the flowfield. Figures 11 and 12 show the density field at three different time instants for the KSPOD and CKSPOD emulations, respectively, for Case A2. Also included are the LES predictions. The KSPOD method is able to predict the rolling vortices upstream of the injector ($x \leq 15$ mm), but produces weakened surface structures. The phenomenon is attributed to the possible difference among the POD modes of the training dataset, which leads to the cancellation of flow dynamics in kriging and reconstruction [12]. The CKSPOD emulation, shown in Fig. 12, on the other hand, can capture most significant rolling vortices and the stringy ligaments revealed in the LES results at all time instants. The traveling surface wave propagates downstream all the way to the injector exit at the speed predicted by the LES method.

To establish broad confidence, the validity of the CKSPOD method is assessed against all test cases. Fig. 13 shows the emulation and simulation results at different time instants. The flow structures and dynamics are well captured. The CKSPOD surrogate model can achieve faithful predictions of the spatiotemporal evolution of the flowfield. Although only the density field is shown here, all other flow variables of interest, such as temperature and pressure, can be emulated using the same procedures based on Algorithm 1.

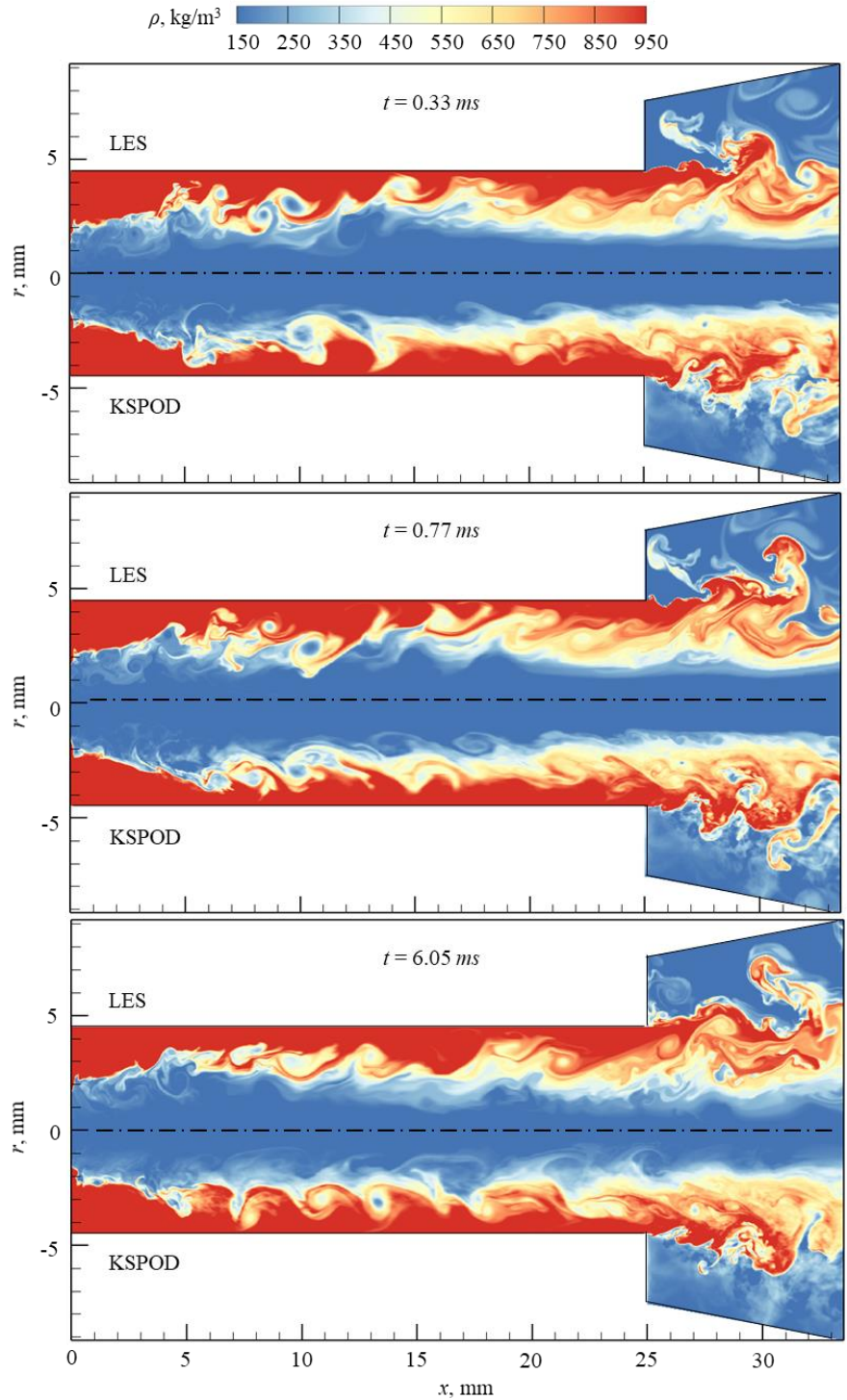

**Fig. 11. Comparison of density field: LES-based simulation and KSPOD emulations. Test Case A2 at three different time instants.**

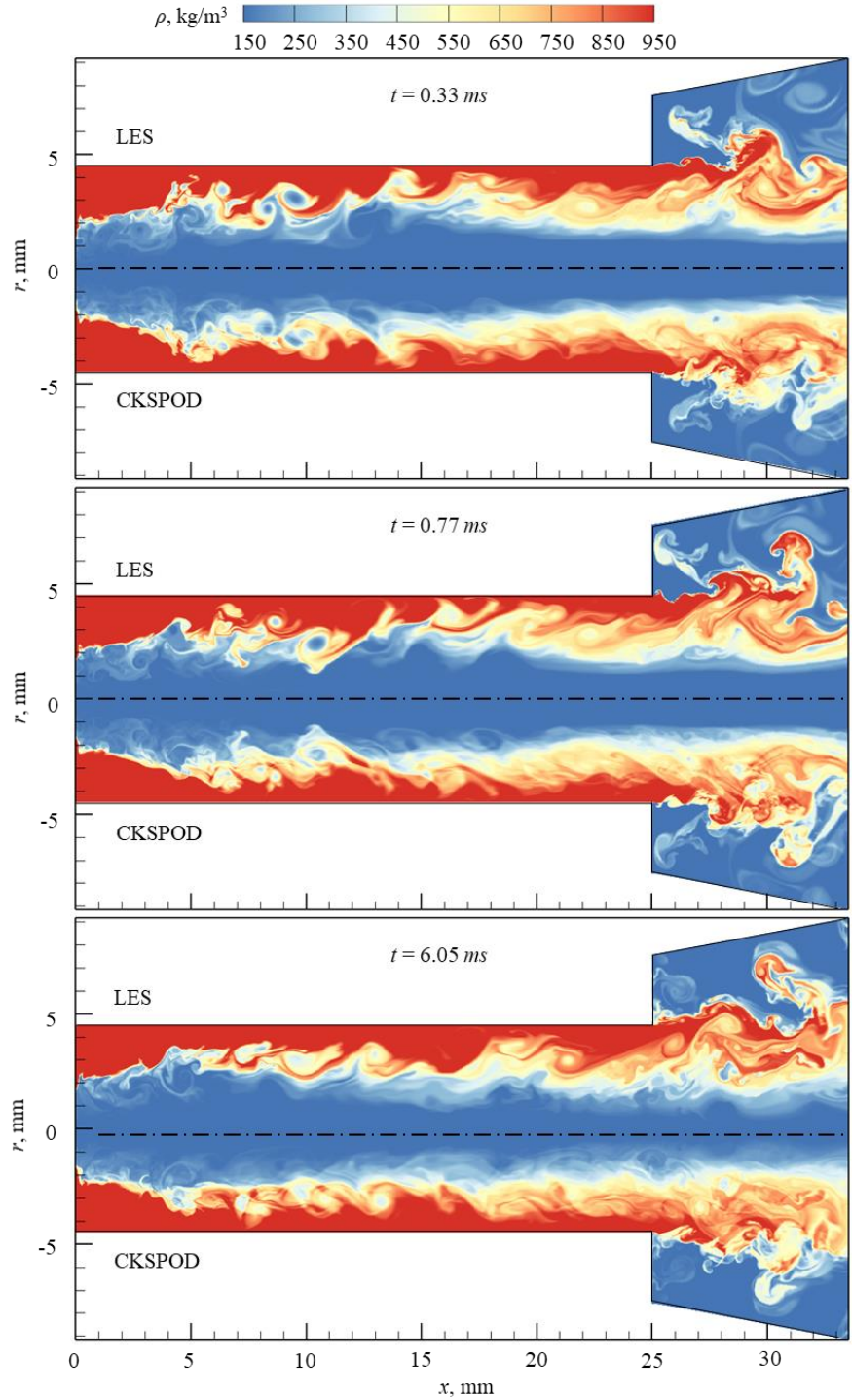

**Fig. 12. Comparison of density field: LES-based simulation and CKSPOD emulations. Test Case A2 at three different time instants.**

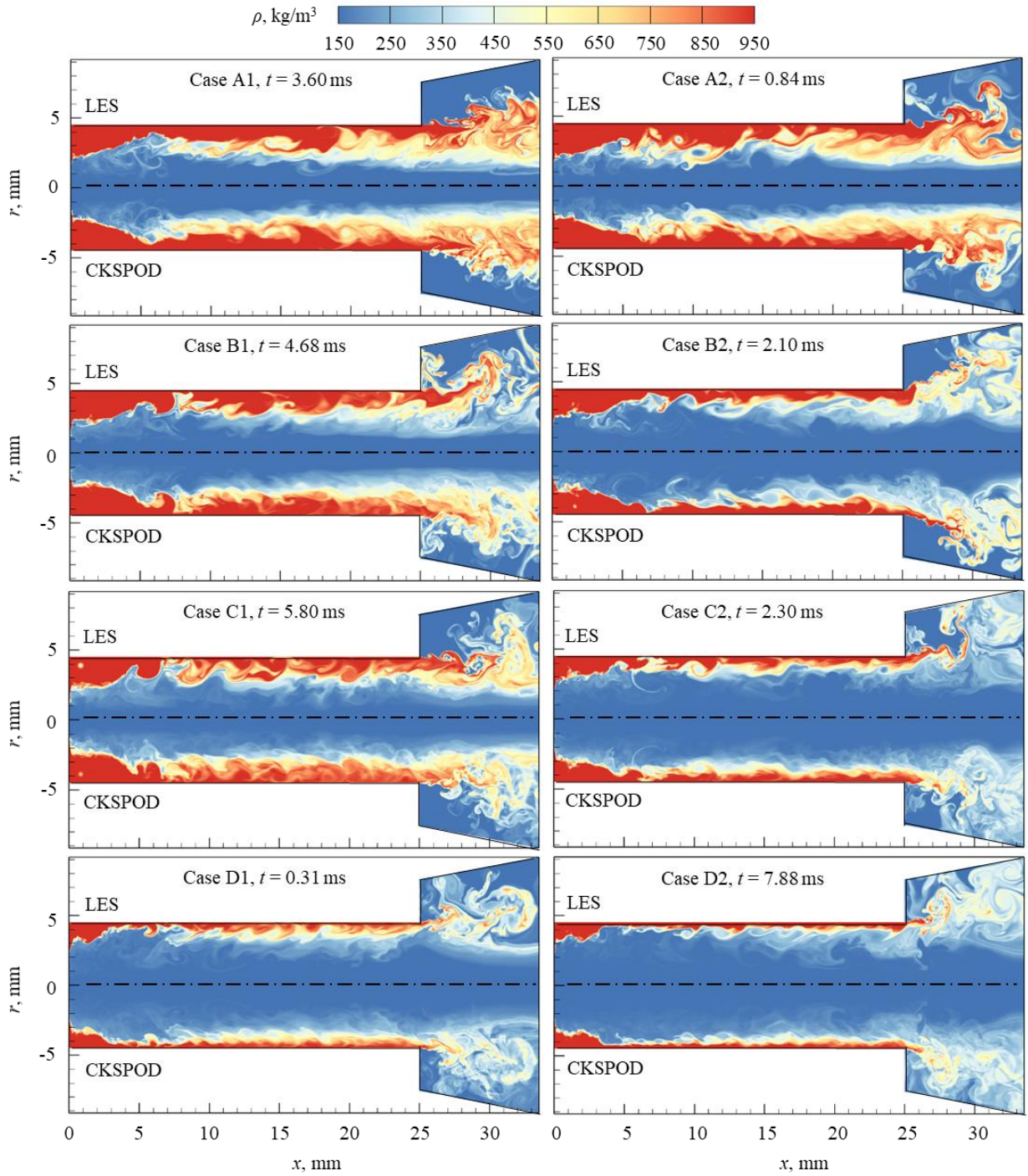

**Fig. 13. Comparison of density fields: LES-based simulation and CKSPOD emulation.**

### 4.5. Performance measures

Two performance metrics, liquid film thickness and spreading angle at the injector exit, are calculated to quantitatively assess the model accuracy. Table 5 lists the time-mean values, standard deviations, and relative errors obtained from the LES, KSPOD, and CKSPOD methods for all validation cases. Time-mean values are determined by averaging the instantaneous data over a statistically meaningful duration. SD denotes standard deviation, and the error is calculated as follows.

$$\varepsilon_r = \frac{|x_{sim} - x_{emu}|}{x_{sim}} \times 100\% \qquad (23)$$

where $x_{sim}$ represents data from simulation and $x_{emu}$ from emulation. Both the KSPOD and CKSPOD emulators are capable of producing high-fidelity results of time-mean flow quantities. The CKSPOD method, however, outperforms the KSPOD method, with errors on the order of 0.1% for most cases.

**Table 5. Time-mean liquid film thickness and spreading angle from simulation and emulation results**

| Case Number | Analysis | Spreading Angle (°) | | | Film Thickness (mm) | | |
|---|---|---|---|---|---|---|---|
| | | LSE | KSPOD | CKSPOD | LES | KSPOD | CKSPOD |
| A1 | Average | 52.846 | 52.919 | 52.857 | 0.629 | 0.625 | 0.628 |
| | SD | 5.185 | 4.976 | 4.392 | 0.169 | 0.162 | 0.136 |
| | Error | - | 0.14% | 0.02% | - | 0.51% | 0.10% |
| A2 | Average | 52.566 | 51.959 | 52.657 | 0.637 | 0.657 | 0.640 |
| | SD | 5.028 | 5.016 | 5.897 | 0.165 | 0.166 | 0.144 |
| | Error | - | 1.15% | 0.17% | - | 3.14% | 0.41% |
| B1 | Average | 54.216 | 53.660 | 54.373 | 0.582 | 0.600 | 0.595 |
| | SD | 4.542 | 4.546 | 4.969 | 0.145 | 0.146 | 0.119 |
| | Error | - | 1.02% | 0.29% | - | 3.03% | 2.25% |
| B2 | Average | 53.811 | 53.875 | 53.819 | 0.594 | 0.592 | 0.594 |
| | SD | 4.226 | 4.130 | 3.732 | 0.136 | 0.132 | 0.111 |
| | Error | - | 0.12% | 0.02% | - | 0.40% | 0.04% |
| C1 | Average | 57.684 | 57.713 | 57.758 | 0.474 | 0.473 | 0.475 |
| | SD | 3.415 | 3.086 | 3.800 | 0.100 | 0.089 | 0.112 |
| | Error | - | 0.05% | 0.13% | - | 0.36% | 0.04% |
| C2 | Average | 57.778 | 57.741 | 57.750 | 0.471 | 0.472 | 0.471 |
| | SD | 3.177 | 3.016 | 3.244 | 0.093 | 0.087 | 0.077 |
| | Error | - | 0.06% | 0.05% | - | 0.13% | 0.02% |
| D1 | Average | 58.998 | 58.031 | 58.786 | 0.379 | 0.379 | 0.379 |
| | SD | 5.389 | 5.146 | 4.860 | 0.107 | 0.105 | 0.120 |
| | Error | - | 1.64% | 0.36% | - | 0.02% | 0.10% |
| D2 | Average | 61.586 | 61.334 | 61.541 | 0.370 | 0.377 | 0.371 |
| | SD | 3.617 | 3.893 | 3.289 | 0.094 | 0.101 | 0.083 |
| | Error | - | 0.41% | 0.07% | - | 1.97% | 0.26% |

The spatial distributions of time-mean liquid film surface along the axial direction are also obtained by averaging more than 1000 snapshots. Figure 14 shows the results along with the LES data. The CKSPOD-predicted liquid film surface closely coincides with the LES predictions, while the KSPOD results show discrepancies near the transient region in Cases B1, C1, and D1.

Figure 15 shows the relative errors of the CKSPOD and KSPOD results. The horizontal dashed lines represent averaged relative error $\bar{\varepsilon}_r$ for each validation case. The CKSPOD emulation consistently outperforms the KSPOD-method. The former has an averaged error of less than 3% for all cases, in comparison with up to 6.4% for the latter. The peak error occurs in the transition stage of the liquid film development and decreases when the film is fully developed near the injector exit. For both the KSPOD and CKSPOD methods, the prediction accuracy near the injector exit increases with increasing inlet velocity magnitude increases from Group A to Group D.

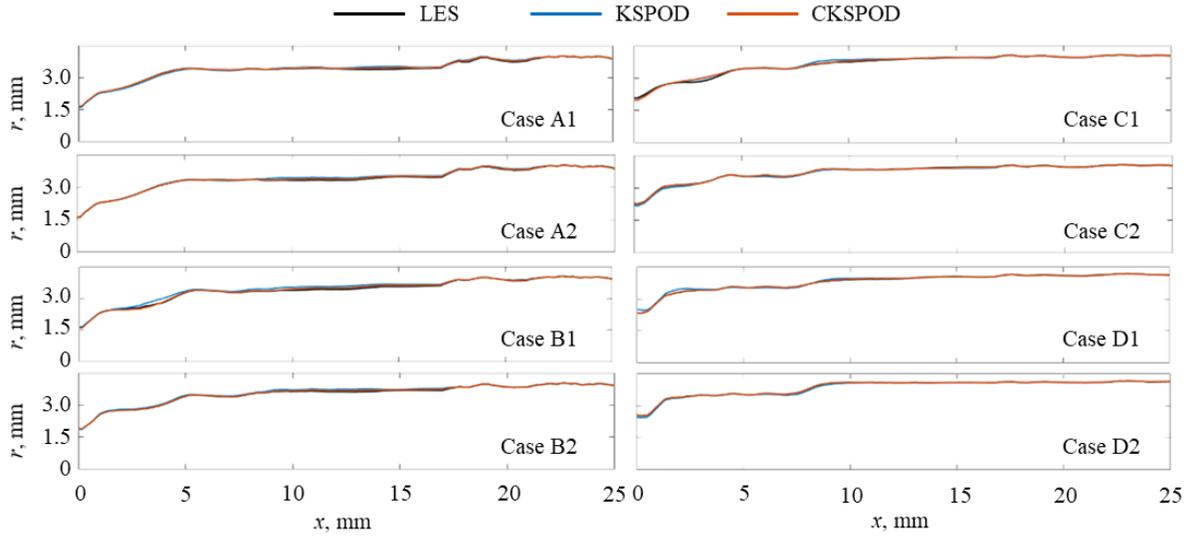

**Fig. 14. Time-mean development of liquid film surface along axial direction, averaged over 1000 snapshots**

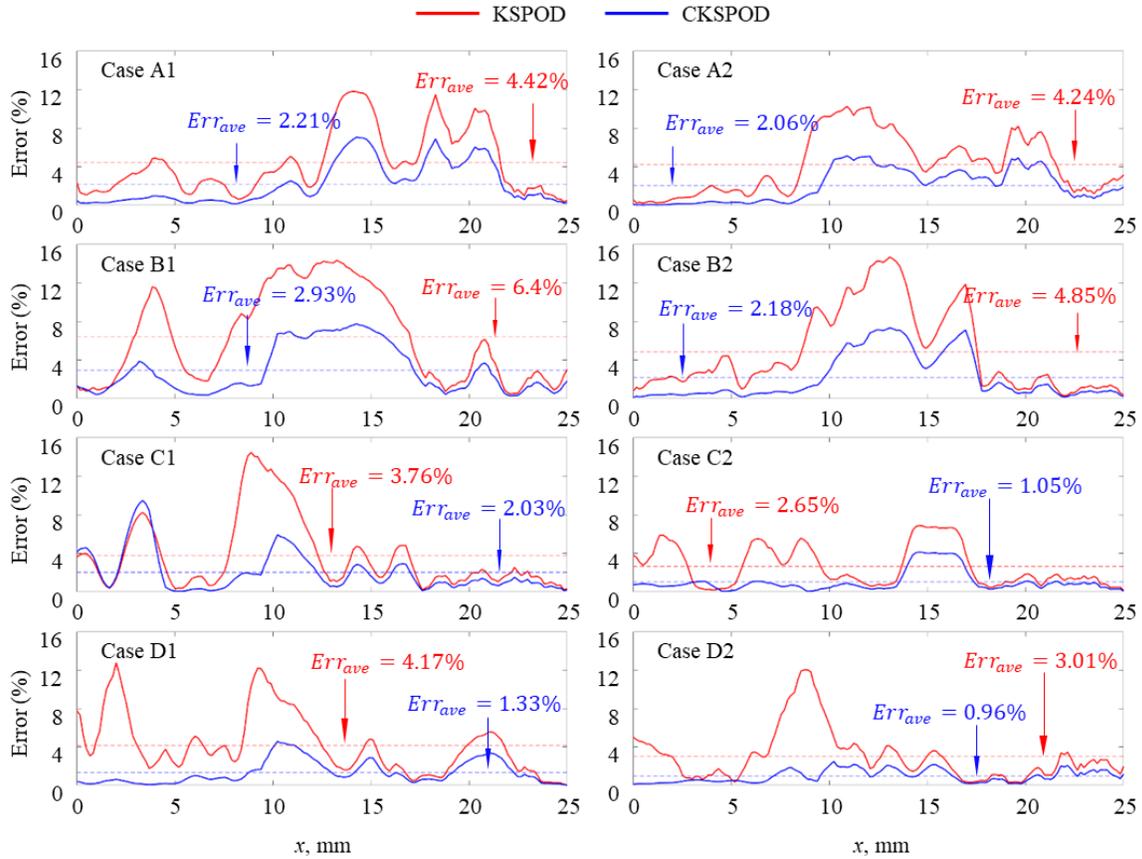

**Fig. 15. Relative error of time-mean liquid film thickness along the axial direction.**

### 4.6. Uncertainty quantification (UQ)

For computer experiments, quantification of prediction uncertainty is as important as the prediction itself. In our earlier study, the one-sided width of the 80% pointwise confidence interval of axial velocity and turbulent kinetic energy was used as a measure for the spatial uncertainty quantification [38]. The work demonstrated the usefulness of uncertainty quantification not only as a measure of predictive uncertainty but also as a means for extracting useful flow physics without expensive simulations. In this subsection, the uncertainty quantification of both KSPOD and CKSPOD emulations are explored. The turbulence kinetic energy defined in Sec. 3 is analyzed.

Figures 16 shows the spatial distribution of time-averaged turbulent kinetic energy predicted by the LES, CKSPOD, and KSPOD models for Case A1 (low inlet velocity) and Case C1 (intermediate high inlet velocity). Higher turbulent kinetic energy occurs along the centerline downstream of the injector exit, where a recirculating flow is formed due to vortex breakdown [1,4]. The turbulent kinetic energy predicted by the CKSPOD method bears close resemblance to that predicted by the LES method. A very similar recirculating flow in the center is shown between CKSPOD and LES cases. The KSPOD emulation leads to considerable over-prediction for Cases A1 and C1. The situation is further corroborated by the standard deviation, as shown in Fig. 17. The smaller the standard deviation, the lower the uncertainties. The CKSPOD method outperforms KSPOD, with a much smaller standard deviation in turbulent kinetic energy for both validation cases. Similar observations are made for all other cases and flow quantities (not shown here).

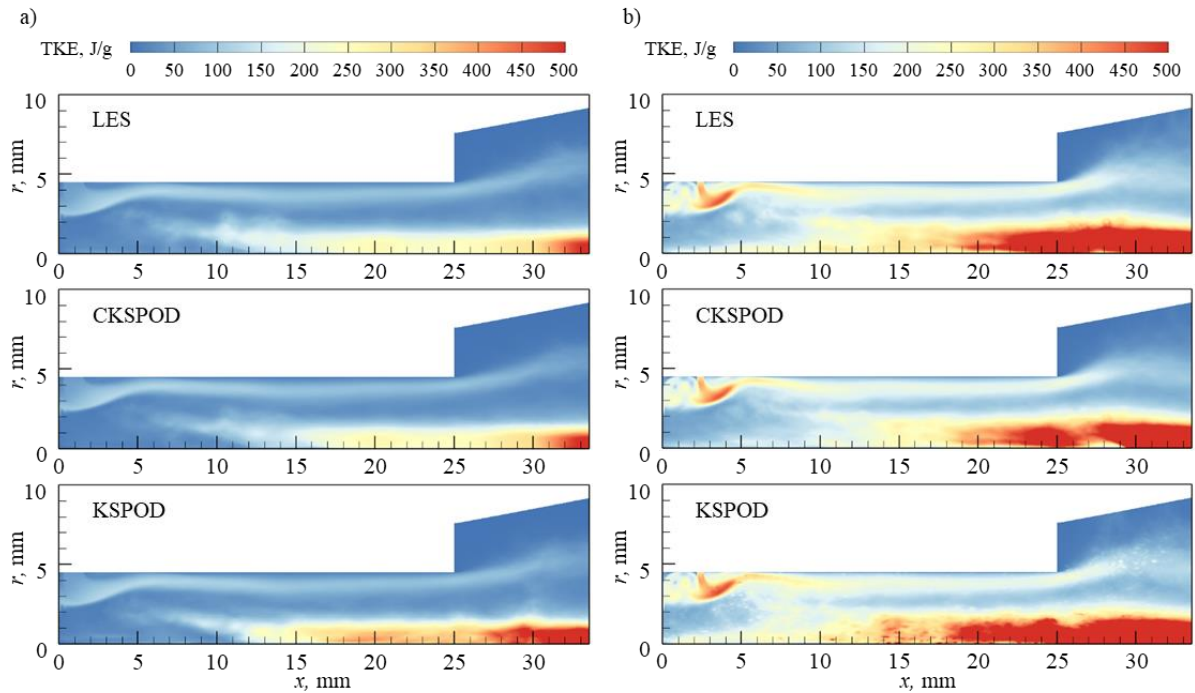

**Fig. 16. Time-averaged turbulent kinetic energy. Cases A1 (left) and C1 (right).**

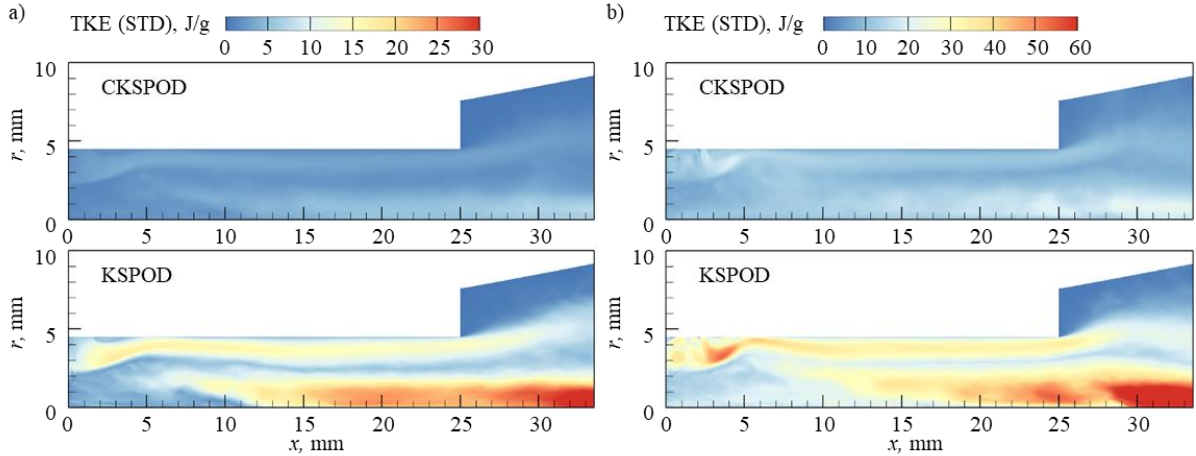

**Fig. 17. Standard deviation of time-averaged turbulent kinetic energy. Cases A1 (left) and C1 (right)**

For Cases A1 and C1, the maximum standard deviation takes place near the LOX inlet ($x = 3.5\ mm$) and the center recirculation ($x = 33\ mm$) downstream of the injector exit. The LOX inlet region contains complex flow structures including separated flows, shear layers, and corner recirculating flows [17,20]. The flow evolution is more dynamic and thus more difficult to predict accurately. This is consistent with the earlier result in Fig. 14 that the first local maximum of error for the liquid-film surface occurs in the LOX inlet area. Similarly, strong vortical dynamics downstream of the injector exit increases the uncertainty of prediction.

### 4.7. Computing time

For the present LES-based simulations, the computing time for different design settings varied in the range of 250-350 CPU hours (hexa-core AMD Opteron Processor 8431). In comparison, the time required to build the surrogate model is about 75 minutes on 10 CPUs, and the time of emulating a new case using the developed surrogate model is less than 10 minutes on 5 CPUs. Therefore, the time saving of emulation is more than 5 orders of magnitude, compared to simulation. The difference between the KSPOD and CKSPOD methods lies in the execution of eigendecomposition; the later takes a slightly more CPU time due to extra procedures for building the common Gram matrix and transfer functions. The overall wall time for the CKSPOD emulator is about 50 seconds of CPU time per snapshot, roughly 1.2 times longer than the KSPOD emulation.

## 5. Conclusion

A new surrogate model based on the common kernel-smoothed proper-orthogonal-decomposition (CKSPOD) technique is proposed for efficient emulation of spatiotemporally evolving flow dynamics. The model requires the construction of a common Gram matrix using the Hadamard product and a transfer matrix, through which all POD modes and time-varying coefficients at each design setting are transferred to the same phase (i.e., no sign differences of eigenvectors among all the spatial modes). The resultant spatial modes and coefficients circumvent the phase-difference issue associated with the kernel-smoothed proper-orthogonal-decomposition (KSPOD) technique. The work is validated against the spatiotemporal flow evolution in a simplex swirl injector with three design parameters. A total of 30 training design settings are selected through the Sliced Latin hypercube design approach. Eight validation cases are considered. Large eddy simulations (LES) are performed at both training and validation settings. For comparison, the emulation results from the KSPOD and CKSPOD methods are presented along with the LES data. The CKSPOD method provides much better predictions overall than the KSPOD counterpart, in terms of time-mean flow quantities and spatiotemporal evolution of the flowfield.

The CKSPOD surrogate model can significantly reduce the computing time at a new design setting by five orders of magnitude, compared to an LES-based simulation. The surrogate model developed in the present work can be effectively applied to a wide range of engineering and scientific problems involving spatiotemporal evolution.

**Acknowledgement**

This work was partly sponsored by the William R. T. Oakes Endowment of the Georgia Institute of Technology. The advice of Dr. Chih-Li Sung and Dr. Shiang-Ting Yeh is gratefully acknowledged.